\input psfig.sty
\magnification=\magstep0
\hsize=13.5 cm               
\vsize=19.0 cm               
\baselineskip=12 pt plus 1 pt minus 1 pt  
\parindent=0.5 cm  
\hoffset=1.3 cm      
\voffset=2.5 cm      
\font\twelvebf=cmbx10 at 12truept 
\font\twelverm=cmr10 at 12truept 
\overfullrule=0pt
\nopagenumbers    
%
\newtoks\leftheadline \leftheadline={\hfill {\eightit Authors' name}
\hfill}
\newtoks\rightheadline \rightheadline={\hfill {\eightit the running title}
 \hfill}
\newtoks\firstheadline \firstheadline={{\eightrm Bull. Astron. Soc. India
(2000)
{\eightbf 28,} } \hfill}
\def\makeheadline{\vbox to 0pt{\vskip -22.5pt
\line{\vbox to 8.5 pt{}\ifnum\pageno=1\the\firstheadline\else%
\ifodd\pageno\the\rightheadline\else%
\the\leftheadline\fi\fi}\vss}\nointerlineskip}
%
\font\eightrm=cmr8  \font\eighti=cmmi8  \font\eightsy=cmsy8
\font\eightbf=cmbx8 \font\eighttt=cmtt8 \font\eightit=cmti8
\font\eightsl=cmsl8
\font\sixrm=cmr6    \font\sixi=cmmi6    \font\sixsy=cmsy6
\font\sixbf=cmbx6 
\def\eightpoint{\def\rm{\fam0\eightrm}
\textfont0=\eightrm \scriptfont0=\sixrm \scriptscriptfont0=\fiverm
\textfont1=\eighti  \scriptfont1=\sixi  \scriptscriptfont1=\fivei
\textfont2=\eightsy \scriptfont2=\sixsy \scriptscriptfont2=\fivesy
\textfont3=\tenex   \scriptfont3=\tenex \scriptscriptfont3=\tenex
\textfont\itfam=\eightit  \def\it{\fam\itfam\eightit}%
\textfont\slfam=\eightsl  \def\sl{\fam\slfam\eightsl}%
\textfont\ttfam=\eighttt  \def\tt{\fam\ttfam\eighttt}%
\textfont\bffam=\eightbf  \scriptfont\bffam=\sixbf
\scriptscriptfont\bffam=\fivebf \def\bf{\fam\bffam\eightbf}%
\normalbaselineskip=10pt plus 0.1 pt minus 0.1 pt
\normalbaselines
\abovedisplayskip=10pt plus 2.4pt minus 7pt
\belowdisplayskip=10pt plus 2.4pt minus 7pt
\belowdisplayshortskip=5.6pt plus 2.4pt minus 3.2pt \rm}
%
%
\def\leftdisplay#1\eqno#2$${\line{\indent\indent\indent%
$\displaystyle{#1}$\hfil #2}$$} \everydisplay{\leftdisplay}
%
\def\frac#1#2{{#1\over#2}}


%
%
\def\pmb#1{\setbox0=\hbox{$#1$}\kern-0.015em\copy0\kern-\wd0%
\kern0.03em\copy0\kern-\wd0\kern-0.015em\raise0.03em\box0}
%
\vglue 50 pt  
%
\leftline{\twelvebf Perspective of long baseline optical interferometry}
%
\smallskip
\vskip 40 pt  
\leftline{\twelverm S. K. Saha\footnote{$^1$}{\eightit e-mail: 
sks@iiap.ernet.in\ \  $^2$ e-mail: smorel@cfa.harvard.edu }, 
and S. Morel$^2$}
\leftline{\eightit $^1$Indian Institute of Astrophysics, Bangalore 560 034,
India. 
}
\leftline{\eightit $^2$Infrared Optical Telescope Array, F. L. Whipple
Observatory, 670 Mt-Hopkins Road, }
\leftline{\eightit Amado AZ 85645, USA. 
}
\bigskip
\noindent{\eightrm Received 10. 1. 2000 ; Accepted 21. 2. 2000}
%
\vskip 20 pt 
%
%
\leftheadline={\hfill {\eightit Saha and Morel} \hfill} 
\rightheadline={\hfill {\eightit Perspective of long baseline optical 
interferometry} \hfill}  
 {\parindent=0cm\leftskip=1.5 cm 
 
{\bf Abstract.} 
\noindent  
This article is a sort of sequel of the earlier extensive review by 
Saha (1999a) where emphasis was laid down on the ground based single aperture, 
as well as  on the working long baseline optical interferometers (LBI) situated 
at the various observatories across the globe that are producing a large amount 
of astronomical results. Since the future of high resolution astronomy 
lies with the new generation of arrays, the numerous technical challenges of  
developing such systems are addressed indicating the current trends and the  
path to future progress in interferometry. The new generation interferometers  
such as Palomar testbed interferometer (PTI), Navy prototype optical 
interferometer (NPOI), Keck interferometer, Very large telescope interferometer  
(VLTI), Center for high angular resolution astronomy (CHARA) array,  
Optical very large array (OVLA), Mitaka optical infrared arrays (MIRA), etc.,  
are being developed. A few of them, viz., PTI, NPOI, IOTA are producing  
results. Among the working interferometers that have been described earlier  
by Saha (1999a), the expansion of the Grand interf\'erom\`etre \`a deux (two)  
t\'elescopes (GI2T), Infrared and optical telescope array (IOTA) are in  
progress. The current status of all these interferometers stated above are  
enumerated. The data analysis being carried out using the working 
interferometers are also described. The space interferometry programmes are 
advancing very fast. Among the notable ones are the Space technology 3 (ST3), 
Space interferometry mission (SIM), and Darwin; they have already received 
funds. The technical details of these interferometers and their objectives are 
highlighted. 
 
}                                 
%
\vfill\eject 
%
%
\vskip 20 pt 
\centerline{\bf Table of contents} 
\bigskip 
\noindent 
\item{1.} Introduction 
\item{2.} Historical development of ground-based interferometry 
 
\indent{2.1.} Intensity interferometry 
 
\indent{2.2.} Single aperture interferometry 
 
\indent{2.3.} Amplitude and phase interferometry 
\item{3.} On-going ground-based interferometric projects 
 
\indent{3.1.} Palomar testbed interferometer (PTI) 
 
\indent{3.2.} Navy prototype optical interferometer (NPOI) 
 
\indent{3.3.} Keck interferometer 
 
\indent{3.4.} Very large telescopes interferometer (VLTI) 
 
\indent{3.5.} Center for high angular resolution astronomy (CHARA) arrays 
 
\indent{3.6.} Grand interf\'erom\`etre \`a deux t\'elescopes (GI2T) current  
status 
 
\indent{3.7.} Infrared-optical telescope array (IOTA) current status 
 
\indent{3.8.} Optical very large array (OVLA) 
 
\indent{3.9.} Mitaka optical infrared arrays (MIRA) 
\item{4.} Data acquisition and processing in optical interferometry 
 
\indent {4.1.} Fringe acquisition and tracking 
 
\indent {4.2.} Data reduction 
 
\item{5.} Image reconstruction 
\item{6.} Space-borne interferometers 
 
\indent{6.1.} Astrometry from space 
 
\indent{6.2.} First space-borne interferometers 
 
\indent{6.3.} Searching for life on other planets 
 
\indent{6.4.} Long-term perspective 
\item{7.} Conclusions 
\item{}Acknowledgments 
\item{}References 
\vskip 20 pt 
\centerline{\bf 1. Introduction} 
\bigskip 
\noindent  
The implementation of imaging by interferometry in optical astronomy 
is a challenging task. Though interferometry at optical wavelengths in 
astronomy began more than a century and a quarter ago (Fizeau, 1868), the 
progress in achieving high angular resolution has been modest. The first 
successful measurement of the angular diameter of $\alpha$~Orionis was 
performed in 1920 using stellar interferometer (Michelson and Pease, 1921), but the field 
lay dormant until it was revitalized by the development of intensity 
interferometry (Brown and Twiss, 1958). Over the last few decades, a marked 
progress has been  witnessed in the development of this field, offering to 
realize the potential of the interferometric technique. 
\bigskip  
Single aperture speckle interferometry (Labeyrie, 1970) decodes the 
diffraction-limited spatial Fourier spectrum and image features of the object. 
A profound increase has been noticed in its contribution (Saha, 1999a and 
references therein) to measure fundamental stellar parameters, viz., (i) 
diameter of stars, (ii) separation of close  binary stars, (iii) imaging of 
emission line of the active galactic nuclei (AGN), (iv) the spatial 
distribution of circumstellar matter surrounding objects, (v) the 
gravitationally lensed QSO's, etc., (Saha, 1999a, 1999b and references therein). 
\bigskip  
Significant improvements in technological innovation over the past several 
years have brought the hardware to compensate in real-time for telescope 
image degradation induced by the atmospheric turbulence that distorts the 
characteristics of light traveling through it. The limitation is due to warping 
of iso-phase surfaces and intensity variation across the wavefront, thereby, 
distorting the shapes of the wavefront (Fried, 1966). The blurring suffered by 
such images is modeled as convolution with the point spread function (PSF). 
Wavefront sensing and adaptive optics (AO) are based on this hardware oriented 
correction (Babcock, 1953, Rousset et al., 1990). 
\bigskip  
Success in synthesizing images obtained from a pair of independent telescopes  
on a North-South baseline configuration (Labeyrie, 1975, Labeyrie et 
al., 1986, Shao et al., 1988), impelled astronomers to venture towards 
ground-based very large arrays (Davis et al., 1992). Potentials for progress 
in the direction of developing large interferometric arrays of telescopes 
(Labeyrie, 1996) are expected to provide images, spectra of quasar host 
galaxies, exo-planets that may be associated with stars outside the solar 
system (Labeyrie, 1995, 1998a, 1998b). Plans are also on to put an 
interferometer of a similar kind on the surface of the moon at the fall of this 
century. The technique of developing long baseline Fizeau-type interferometer 
for lunar operation consisting of 20 to 27 off-axis parabolic segments carried 
by robotic hexapodes that are movable during observing run has been suggested by 
Arnold et al., (1996). In a very recent article, Saha (1999a) has discussed at 
length about the interferometric techniques, that include the basic features of 
the working long baseline interferometers (LBI) with two or more optical 
telescopes. This review focuses on the current activities of the various groups 
across the globe to develop new ground-based, as well as space-borne 
interferometers in the optical domain, data processing techniques being adapted 
at the working LBIs; an account of historical development of high resolution 
astronomy is enunciated for the benefit of the readers as well. Some of the 
important results obtained with the new interferometers are also highlighted. 
\vskip 20 pt 
\centerline{\bf 2. Historical development of ground-based interferometry} 
\bigskip 
\noindent    
High angular resolution of an stellar object is an important aspect 
which astronomers are aspiring for. Ever since Fizeau (1868) had suggested to 
install a screen with two holes on top of the telescope that produce Young's 
fringes at its focal plane as the fringes remain visible in presence of seeing, 
several attempts have been made with moderate sized telescopes to measure 
stellar diameters. St\'ephan (1874) tried to resolve Sirius by using several 
masks with hole separation up to 65 cm on the 80 cm telescope of 
Observatoire de Marseille (France). No fringe contrast change was noticed and,  
therefore, only a maximum diameter of Sirius was deduced. One of the first  
significant results was the measurement of diameter of the satellites of  
Jupiter with a Fizeau interferometer on top of the Yerkes refractor by  
Michelson (1891). With the 100 inch telescope at Mt. Wilson 
(Anderson, 1920), the angular separation of spectroscopic binary star Capella 
was also determined. 
\bigskip  
To overcome the restrictions of the baseline, Michelson (1920) constructed the 
stellar interferometer equipped with 4 flat mirrors to fold the beams by 
installing a 7~m steel beam on top of the telescope afore-mentioned 
100 inch telescope; the supergiant star $\alpha$~Orionis were resolved 
(Michelson and Pease, 1921). Due to the various difficulties, viz., (i) effect 
of atmospheric turbulence, (ii) variations of refractive index above small 
sub-apertures of the interferometer, and (iii) mechanical instability, the 
project was abandoned. 
\vskip 20 pt 
\noindent   {\bf 2.1. Intensity interferometry} 
\bigskip 
\noindent    
The field of optical interferometry lay dormant until it was 
revitalized by the development of intensity interferometry (Brown and Twiss, 
1958). Success in completing the intensity interferometer at radio wavelengths 
(Brown et al., 1952), in which the signals at the antennae are detected 
separately and the angular diameter of the source is obtained by measuring 
correlation of the intensity fluctuations of the signals as a function of 
antenna separation, Brown and Twiss (1958) demonstrated its potential at optical 
wavelengths by measuring the angular diameter of Sirius. Subsequent development 
of this interferometer with a pair of 6.5 meter light collector on a circular 
railway track spanning 188 meter (Brown et al., 1967), depicted the measurements 
of 32 southern binary stars with angular resolution limit of 0.5 milliarcseconds 
(Brown, 1974). The project was abandoned due to lack of photons beyond 2.5 
magnitude stars. 
\vskip 20 pt 
\noindent   {\bf 2.2. Single aperture interferometry} 
\bigskip 
\noindent    
Meanwhile, Labeyrie (1970) had invented speckle interferometric 
technique that retrieves the diffraction-limited information of an object. The 
diffraction-limited resolution of celestial objects viewed through the  Earth's 
turbulent atmosphere could be achieved with the large optical  telescope, by 
post detection processing of a large data set of short-exposure  images using 
Fourier-domain methods. Certain specialized moments of the Fourier transform of 
a short-exposure image contain diffraction-limited information  about the 
object 
of interest. Owing to the turbulent phenomena associated  with heat flow and 
winds in the atmosphere, the density of air fluctuates in  space and time. The 
inhomogeneities of the refractive index of the air can  have devastating effect 
on the resolution achieved by any large telescope. The  disturbance takes the 
form of distortion of the shape of the wavefront and  variations of the 
intensity across the wavefront. Due to the motion and temperature fluctuations 
in the air above the telescope aperture, inhomogeneities in the refractive 
index 
develop. These inhomogeneities have the effect of breaking the aperture into 
cells with different values of refractive index that are moved by the wind 
across the telescope aperture. 
\bigskip  
The power spectral density of refractive index fluctuations caused by 
the atmospheric turbulence follows a power law with large eddies having greater 
power (Tatarski, 1967). A plane wave propagating through the atmosphere of 
Earth is distorted by refractive index variation in the atmosphere 
(troposphere); it suffers phase fluctuations and reaches the entrance pupil of 
with patches of random excursions in phase (Fried, 1966). 
Therefore, the image of the star in the focal plane of a telescope 
is larger than its Airy disk (theoretical size 1.22$\lambda$/D is known as
Rayleigh limit or diffraction limit, where, D is the diameter of the 
telescope). The size is equivalent to the atmospheric point spread function 
(point spread function is a modulus square of the Fourier transform of the 
aperture function). The resolution at the image plane of the telescope is 
determined by the width of the PSF which is of the order of 
(1.22$\lambda/r_0$), where, $\lambda$ is a wavelength of light and $r_0$ is 
the average size of the turbulence cell, which is of the  order of 10~cm. 
The statistical properties of speckle pattern depend both on the 
coherence of the incident light and the properties of random medium. 
Mathematically, speckles are simply the result of adding  many sine functions 
having different, random characteristics. Since the  positive and negative 
values cannot cancel out everywhere, adding an infinite  number of such sine 
functions would result in a function with 100 $\%$  constructed oscillations. 
Further details about the technique, way of recording speckles of any 
astronomical objects, image processing can be found in the most recent article 
by Saha (1999a, 1999b). 
\bigskip  
The afore-mentioned technique has been successful in obtaining 
spectacular results of a wide range of objects; a glimpse of such studies  can 
be found in a recent review by Saha (1999a and references therein).  Studies of 
the morphology of stellar atmospheres, the circumstellar environment of nova or 
supernova, YPN, long period variables (LPV), rapid variability of AGNs etc., 
are of paramount importance in astrophysics. Details of the structure of a wide 
range of stellar objects at scales of  
0.015$^{\prime\prime}$~-~0.03$^{\prime\prime}$ are routinely observed. The 
physical properties of red dwarfs in the vicinity of sun can also be looked 
into; some dwarfs may often be close binaries. Speckle interferometric 
technique has been extended to IR domain too. With the photon counting detector 
system (Saha, 1999a and references therein) which is an essential tool in the 
application of optical interferometric imaging that allows the accurate photon 
centroiding, as well as provides dynamic range needed for measurements of source 
characteristics, one can record the specklegrams of the object of faintest 
limiting magnitude. Further benefits have been witnessed when the 
atmospherically degraded images of these objects are applied to image 
restoration techniques (Liu and Lohmann, 1973, Rhodes and Goodman, 1973, Knox 
and Thomson, 1974, Lynds et al., 1976, Weigelt, 1977, Lohmann et al., 1983, 
Ayers and Dainty, 1988) for obtaining Fourier phase. Mapping the finer features 
of such objects would produce qualitative scientific results. 
\bigskip  
Development of various interferometric techniques, namely, (i) speckle 
spectroscopy (Grieger and Weigelt, 1992), (ii) speckle polarimetry (Falcke et 
al., 1996), (iii) pupil plane interferometry (Roddier and Roddier, 1988), (iv) 
Closure-phase method (Baldwin et al., 1986), (v) aperture synthesis using both 
partial redundant and non-redundant masking (Haniff et al., 1987, 1989, 
Nakajima et al., 1989, Busher et al., 1990, Bedding et al., 1992, 1994, Bedding, 
1999), (vi) differential speckle interferometry (Petrov et al., 1986) too 
contribute in 
obtaining new results. Adaptive optics system introduces  controllable counter 
wavefront distortion which both spatially and temporally follows that of the 
atmosphere. Adaptive optical systems may become standard  tool for the new 
generation large telescopes. A considerable amount of new results have already 
been published. Detailed technique and the results can be seen in the recent 
reviews (L\'ena, 1997, L\'ena and Lai, 1999a, 1999b, Saha, 1999a). 
\vskip 20 pt 
\noindent   {\bf 2.3. Amplitude and phase interferometry} 
\bigskip 
\noindent    
Subsequently, Labeyrie (1975) had developed a long baseline optical 
interferometer $-$ Interf\'erom\`etre \`a Deux T\'elescopes 
(I2T) $-$ with a pair of 25~cm telescopes at Observatoire de Calern, France, 
exploiting the concept of merging speckles from both the telescopes. In other 
words, the fringed speckle can be visualized when a speckle from one telescope 
is  merged with the speckle from the other telescope. His design combines 
features of the Michelson and of the radio interferometers. The use of 
independent telescopes increases the resolving capabilities. In this case, 
Coud\'e 
beams from both the telescopes arrive at central station and recombines them. 
Apart from the first measurements for a number of giant stars (Labeyrie, 1985), 
this interferometer also determined the effective temperatures of giant stars 
(Faucherre et al., 1983). In addition, resolving the gas envelope of the Be star 
$\gamma$~Cassiopeiae in the H$\alpha$ line (Thom et al., 1986) has a major 
achievement from I2T. In the infrared, diameters of cool bright giants and 
their effective temperature at 2.2~$\mu$m (DiBenedetto and Rabbia, 1987) have 
also been measured. 
\bigskip  
Following the success of its operation, Labeyrie (1978) undertook a 
project of building large interferometer known as Grand 
Interf\'erom\`etre \`a Deux (two) T\'elescopes (GI2T) at the same observatory.  
This interferometer comprises of two 1.5 meter spherical telescopes on a  
North-South baseline, which are movable on a railway track (Labeyrie et al.,  
1986). The first scientific result came out of this interferometer in 1989 
(Mourard et al., 1989) that had resolved the rotating envelope of hot star 
$\gamma$~Cassiopeiae. This object has been the favorite  target to the GI2T 
(Stee et al., 1995, 1998). The emerging results on $\beta$~Lyrae, $\delta$~Cep  
(new and accurate distance estimate), P~Cyg (the first discovery of an  
asymmetry in its wind) and $\zeta$~Tau (the first evidence 
for a one-armed oscillation in a Be star equatorial disk) are the most 
spectacular results from the GI2T too (Mourard et al., 1997, Harmanec et al., 
1996, Vakili et al., 1997, 1998a, 1998b). The technical details of this kind of 
interferometers can be found in the recent article by Saha (1999a). 
\bigskip  
There are several long baseline interferometers, viz., (i) Mt. Wilson 
stellar interferometer, (ii) Sydney University stellar interferometer (SUSI), 
(iii) Cambridge Optical Aperture Synthesis Telescope (COAST), (iv) 
Infrared Optical Telescope Array (IOTA) are in operation. The technical details 
of these interferometers, as well as the results obtained so far can be found 
in the review article by Saha (1999a). However, a few salient objectives and 
programmes of these are reported in brief. 
\bigskip  
Measurements of precise stellar positions and motions of the stars are 
the major programmes being carried out with the Mark III interferometer at Mt. 
Wilson (Shao et al., 1990, Hummel, 1994). This set up has also been used to 
derive the fundamental stellar parameters, like the orbits for spectroscopic, as 
well as eclipsing binaries (Armstrong et al., 1992a, 1992b, Pan et al., 1992, 
Shao and Colavita, 1994), structure of circumstellar shells (Bester et al., 
1991) etc. 
\bigskip  
The interferometers, namely, COAST, SUSI, IOTA etc., are relatively new. The 
expansion of a few of them are in progress. Nevertheless, they have produced 
several spectacular results to be mentioned. Among the notable results 
with COAST, mapping of the double-lined spectroscopic binary 
$\alpha$~Aurigae, resolving of $\alpha$~Tau are the important ones (Baldwin et 
al., 1996, 1998); detecting a circularly symmetric data with an  unusual 
flat-topped and limb darkening profile of $\alpha$~Orionis (Burns et al., 1997), 
variations of the cycle of pulsation  of Mira variable R~Leonis (Burns et al., 
1998) etc., have also been reported. With the SUSI interferometer, Davis et  
al., (1998, 1999) have determined the diameter of $\delta$~CMa with an accuracy 
of $\pm$1.8\%. The results with the IOTA interferometer so far reported are  
from the near IR bands. They are in the form of measuring the angular  
diameters and effective temperatures of carbon stars (Dyck et al., 1996a),  
carbon Miras and S types (Van Belle et al., 1997), K and M giants and  
supergiants (Dyck et al., 1996b, 1998, Perrin et al., 1998), Mira variables  
(Van Belle et al., 1996), and cool giant stars (Dyck et al., 1995). 
\vskip 20 pt 
\centerline{\bf 3. On-going ground-based interferometric projects} 
\bigskip 
\noindent  
In view of the growing importance of high angular resolution 
interferometry, several projects of developing LBIs are in progress. Since the 
technical details of the well established interferometers have been already 
described in the recent article by Saha (1999a), the salient features of some 
of the on-going interferometric projects are enumerated. 
\vskip 20 pt 
\noindent   {\bf 3.1. Palomar testbed interferometer (PTI)} 
\bigskip 
\noindent  
The Palomar testbed interferometer (PTI), is an infrared  
phase-tracking interferometer in operation situated at Palomar Observatory,  
California; it was developed by the 
Jet Propulsion Laboratory and California Institute of Technology for NASA as a 
test-bench for the Keck interferometer. The main thrust of this interferometer 
is to develop techniques and methodologies for doing narrow angle 
astrometry for the purpose of detecting extra-solar planets (Wallace et al., 
1998) that measures the wobble in the position of 
a star caused by the transverse component of a companion's motion. 
\bigskip 
Three 40~cm siderostats (steerable flat mirrors for sending starlight in a fixed 
direction) coupled to beam compressors (reducing the beam diameter) can be used 
pairwise to provide baselines up to 110~m (Colavita et al., 
1999). This interferometer tracks the white light fringes using an array 
detector at 2.2 $\mu$m (K band) and active delay-lines with a range of 
$\pm$38~m. Among others, the notable feature of this interferometer is that of 
implementation of a dual-star astrometric ability; observation of fringes 
from 2 close stars simultaneously for phase referencing and narrow-angle  
astrometry. An end-to-end heterodyne laser metrology system is used to measure 
the optical path length of the starlight (Wallace et al., 1998). They also 
claimed the better performances after the recent  
upgradations of PTI, viz., a single mode fiber for spatial filtering, 
vacuum pipes to relay the beams, accelerometers on the siderostat mirror etc.  
\bigskip  
Malbet et al., (1998) have resolved the young stellar object FU~Orionis 
using the said interferometer in the near infrared with a projected resolution 
better than 2~AU. Observations of the young binary stars, $\iota$~Peg have been 
also conducted by Pan et al., (1996) with this interferometer. They have 
determined its visual orbit with separation of 1~mas in R. A., having a 
circular orbit with a radii of 9.4~mas. Measurements of diameters and effective 
temperatures of G, K, and M giants and  supergiants have been reported recently 
by Van Belle et al., (1999). The visual orbit for the spectroscopic binary 
$\iota$~Peg with interferometric visibility data recoded by PTI has also been 
derived (Boden et al., 1999). 
\vskip 20 pt 
\noindent  {\bf 3.2. Navy prototype optical interferometer (NPOI)} 
\bigskip 
\noindent   
The astrometric array of 
NPOI, a joint project of the US Naval Research laboratory and the US Naval 
Observatory is designed to measure positions with precision comparable to that 
of Hipparcos (1997). This interferometer is located at the Lowell Observatory, 
Arizona and is capable of maintaining accuracy of the positions of the brightest 
Hipparcos stars while improving the precision of their proper motions. The 
anticipated wide-angle astrometric precision of the NPOI is about $\sim$2~mas. 
(Armstrong et al., 1998). Since the high precision astrometry is an important 
aspect to astronomy that helps in establishing cosmic distance scale, 
measurements of proper motion can confirm stars as members of cluster (known 
distance), may elucidate the dynamics of the  Galaxy. NPOI plans to measure the 
positions of some radio stars that would help in matching radio sources with 
their optical counterparts. 
\bigskip  
This interferometer includes sub-arrays for imaging and for astrometry 
and  is developed at Y-shaped (Very Large Array-like) baseline  configuration. 
The light beams are passed through vacuum pipes to the central laboratory. For 
astrometric mode, 4 fixed siderostats (0.4~m diameter) are  used with the 
baselines extendable from 19~m to 38~m (Armstrong et al., 1998). The shared 
back and covers 450-850~nm in 32 channels. The other notable features are  
being the delay 
system, active group-delay fringe tracking etc. The astrometric sub-array has a 
laser metrology system to measure the motions of the siderostats with respect to 
one another and to the bedrock. While for imaging mode, 6 transportable 
siderostats (0.12~m diameter) are used with the baselines from 2~m to 432~m.  
Three siderostat positions are  kept with equal space for each arm of the Y. 
Coherence of imaging configuration is maintained by phase bootstrapping (see  
section 4.1). 
Observations in visible spectrum with 3-elements have been carried out using 
avalanche photo-diode as detector (Hummel et al., 1998). The dynamic range in 
the best of the NPOI images exceeds 100:1 (Armstrong et al., 1998). 
\bigskip  
A few examples of science with NPOI can be read from the following 
results. Pauls et al., (1998) have measured the limb darkened angular diameters 
of late-type giant stars using the said interferometer with three optical 
elements; measurement of non-zero closure phase has been performed on a single 
star. Hajian et al., (1998) have also observed the limb darkened diameters  of 
two K giants, $\alpha$~Arietis and $\alpha$~Cassiopeiae with 20 spectral 
channels covering 520-850~nm. They were able to extend the  spatial frequency 
coverage beyond the first zero of the stellar visibility  function for these 
stars. Hummel et al., (1998) have determined the orbital  parameters of two 
spectroscopic binaries, $\zeta$~Ursae Majoris (Mizar A), 
$\eta$~Pegasi (Matar) and derived masses and luminosities based on the data 
obtained with said interferometer; published radial velocities and Hipparcos 
trigonometrical parallaxes were used for the analysis. 
\vskip 20 pt 
\noindent   {\bf 3.3. Keck interferometer} 
\bigskip 
\noindent   
The development of Keck interferometer consisting of 2 $\times$ 10~m 
apertures (main telescopes) with a fixed baseline of 85~m is in progress: the 
expected resolution of this is of the order of 5~mas at 2.2~$\mu$m. The 
baselines available with outrigger telescopes (4 $\times$ 1.8~m) will be between 
25~m to 140~m (fixed baselines). For imaging the main telescope would be used  
with outriggers (Colavita et al., 1998). This project is  funded by NASA and is 
being carried out by Jet Propulsion Laboratory (JPL) and California Association 
for Research in Astronomy (CARA). This large interferometer is located at Mauna 
Kea Observatory, Hawaii. It will combine phased pupils provided by adaptive  
optics for the main telescopes (up to V=9, 39 mas FWHM, Strehl ratio=30\%) and  
fast tip/tilt correction on the outriggers. Beam recombination will be carried 
out by 5 two-way combiners at 1.5-2.4~$\mu$m for fringe tracking, astrometry,  
and imaging. Project for a 10~$\mu$m  nulling-combiner for exo-zodiacal disk 
characterization is also undertaken. 
\bigskip  
The astrometric accuracy is expected to be of the order of 20 to 30 
$\mu$as/$\sqrt{\rm hour}$. With the main telescopes observations for 
searching Jovian planets, as well as for characterizing the exo-zodiacal disks 
may commence by 2001 and with the addition of the outriggers, astrometric 
observation will be carried out by 2003. 
\vskip 20 pt 
\noindent   {\bf 3.4. Very large telescope interferometer (VLTI)} 
\bigskip 
\noindent 
The VLTI, built by the European Southern Observatory and located at Cerro 
Paranal, Chile, will be a versatile facility consisting of four 8 m 
fixed telescopes (``unit telescopes'') and three 1.8~m mobile auxiliary 
telescopes which can be installed on any of the 30 stations built on the 
ground. The maximum baseline of VLTI is 200~m. Siderostats for first tests 
could be used as well (Derie et al., 2000). Coud\'e beams from these apertures  
are sent through delay-lines 
operating in rooms at atmospheric pressure but at a thoroughly controlled 
temperature in order to avoid turbulence. The beams reach an optical switch-yard 
to be directed to one of the four expected recombiners: VINCI (Kervella et al., 
2000), MIDI (Leinert and Graser, 1998), AMBER (Petrov et al., 2000) or PRIMA 
(Quirrenbach et al., 1998). 
\bigskip 
VINCI (VLT INterferometer Commissioning Instrument) will be a single-mode 
fiber recombiner operating at 2.2 $\mu$m like FLUOR (see 3.7) and is intended to 
be used for debugging the upstream sub-systems of VLTI. MIDI (MID-Infrared) will 
be a beamsplitter-based recombiner that will be used for observations at  
10~$\mu$m. AMBER (Astronomical Multiple BEam Recombiner) has been designed for 
observations between 1~$\mu$m and 2.5~$\mu$m. It will be able to perform 
recombination of three beams in order to use closure-phase techniques. PRIMA 
(Phase-Reference Imaging and Microarcsecond Astrometry) is a recombiner 
dedicated to narrow-angle astrometry (see 6.1). It should be able to reach a 10 
$\mu$as resolution. 
\bigskip 
By the end of 1999, two unit telescopes were operating and the primary mirror  
has 
been installed on the third one. First fringes of the VLTI with two siderostats 
and VINCI are scheduled for the end of 2000. All the remaining combiners are 
scheduled to work either with the unit or the auxiliary telescopes by 2005. 
\vskip 20 pt 
\noindent   {\bf 3.5. Center for high angular resolution astronomy (CHARA) 
array} 
\bigskip 
\noindent 
The Center for High Angular Resolution Astronomy (CHARA), Georgia State  
University, is currently building an interferometric array at Mt. Wilson, 
California, USA; it comprises  
six fixed 1~m telescopes arranged in a Y-shaped configuration with 
a maximum baseline of $\sim$350~m that would operate at optical and IR wave  
band (McAlister et al., 1998) with a limiting resolution of 0.2~mas. The 
main objective of this project is to measure the diameters, distances, masses 
and luminosities of stars, as well as to image features, viz., spots and flares 
on their surfaces. The aim of this project range from detecting other  
planetary systems to imaging the black hole driven central engines of quasars 
and active galaxies. 
\bigskip 
Light from the telescopes is sent through vacuum pipes to the centrally 
located Beam synthesis facility, a L-shaped long building 
(McAlister et al., 1994) that houses 
the optical path length equalizer (OPLE) and the beam combination laboratory 
(BCL). Constructions of piers for the five telescopes (McAlister et al., 1998) 
have been completed. Optical delay line carts and their metrology,   
similar versions that are used at NPOI and PTI, and 
control systems, are being developed at JPL, USA. The first fringes in K band 
from a star have been acquired with two telescopes of CHARA by the end of 1999. 
The other baselines and the visible spectrum recombiner should be operational 
in the next few years. 
\vskip 20 pt 
\noindent  {\bf 3.6. Grand interf\'erom\`etre \`a deux t\'elescopes (GI2T) current status} 
\bigskip 
\noindent 
The detailed description of GI2T is available in a recent article by Saha  
(1999a). This interferometer has recently been upgraded with a new recombiner  
named REGAIN (REcombineur pour GrAnd Interf\'erom\`etre). This recombiner, able 
to operate with three  
telescopes (Mourard et al., 1998) has the following features. 
\bigskip 
The 76 mm Coud\'e beams coming from the telescopes are first compressed to 5 mm 
in order to stabilize the pupil image in a fixed plane. Then, field rotators 
consisting of four plane mirrors are used for each beam to compensate the 
polarization difference affecting the visibility measured (Rousselet-Perraut et 
al., 1996). The different chromatic dispersion between the two beams (due to 
operation at atmospheric pressure) is compensated by using for each beam two 
prisms which can slide on their hypotenuse, forming therefore a plate with 
adjustable thickness. This thickness is modified every 4 minutes, following the 
variation of the altitude of the observed object. Figure 1 depicts the process 
performed by an arm of the REGAIN table prior to recombination. Unlike the  
previous recombiner which had to move to track fringes, REGAIN uses a  
delay-line named LAROCA 
(Ligne A Retard de l'Observatoire de la C\^ote-d'Azur) featuring a cat's eye 
reflector with a variable curvature mirror. An adaptive optics for the 1.5 m 
telescopes of GI2T is in development (V\'erinaud et al., 1998). 
\bigskip 
\midinsert 
{\eightpoint   
\noindent 
\centerline{\psfig{figure=regain.epsi,height=5cm,width=13cm}}  
\bigskip
\noindent 
{\bf Figure 1.} Optical processing of a beam from one telescope of GI2T by the  
REGAIN recombiner.  
} 
\endinsert 
\bigskip 
The focal instrument of GI2T/REGAIN will be a spectrometer working either in 
dispersed fringes mode or in Court\`es mode. In dispersed fringes (see 4.1), 
the spectral range is 480 nm to 750 nm and the spectral resolution can reach 
$R=30,000$. Separated recombination of two orthogonal polarizations will be 
possible. The detectors used will be two CP40 photon cameras. The Court\`es 
mode consists in forming images at different wavelengths of speckles 
with fringes given by the recombination. The spectrometer of REGAIN in 
Court\`es mode will be able to give 16 images at the same time. 
\bigskip 
However, first fringes (at 2.2 $\mu$m) with GI2T/REGAIN have already been 
acquired in August 1999 (Weigelt et al., 2000) thanks to a different slit  
spectrometer and a PICNIC infrared 
camera built by the Max-Planck-Institute f\"ur Radioastronomie (Bonn, Germany). 
\vskip 20 pt 
\noindent   {\bf 3.7. Infrared optical telescope array (IOTA) current status} 
\bigskip 
\noindent    
A description of IOTA has already been presented in this journal 
(Saha, 1999a). IOTA has recently been upgraded (Traub et al. 2000) with a  
supplementary optical path delaying system (consisting of a long-travel  
delay-line fixed while observing and a short-travel delay-line tracking the  
star), a new control system for the long delay-lines and 
new high-precision secondary mirror holders for the telescopes. The third 
collector similar to the two already existing (a 45 cm siderostat and a fixed 
Mersenne telescope compressing the beam by 10) will soon be  operational. 
Identical reflections are applied to the beams from the collectors to the  
recombiner in order to avoid any loss of fringe contrast due to different  
polarization states between the two beams at the recombination point. Figure 2  
depicts the overview of IOTA interferometer. 
\bigskip 
\noindent 
\midinsert 
{\eightpoint  
\noindent 
\centerline{\psfig{figure=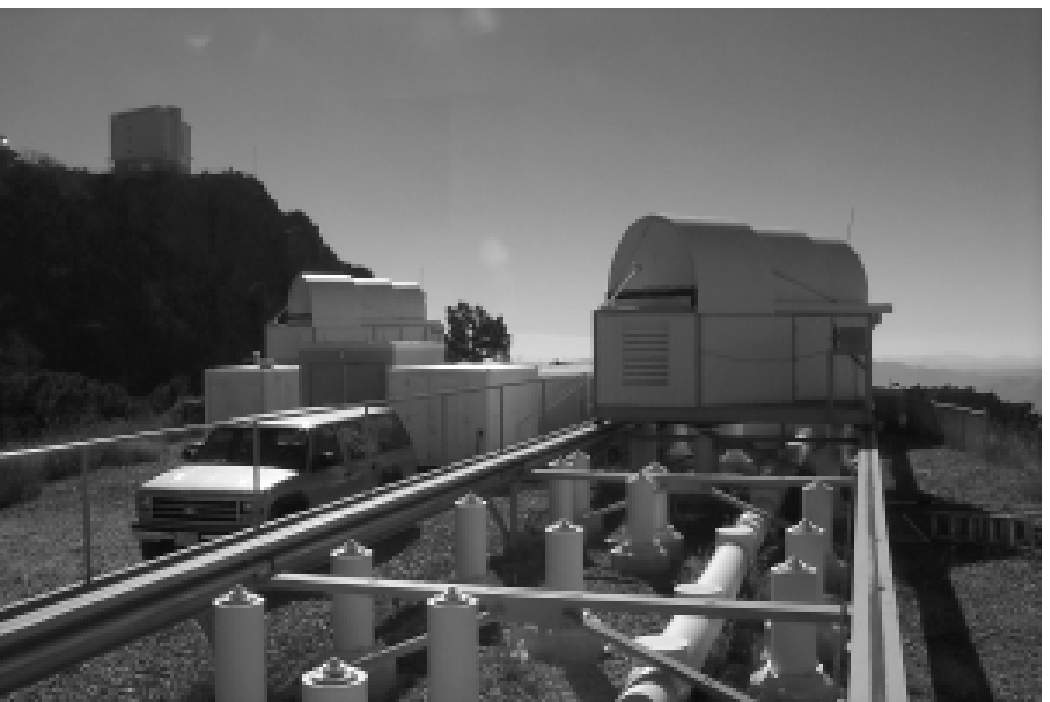,height=9cm,width=11cm}}  
\bigskip 
\noindent 
{\bf Figure 2.} Overview of IOTA interferometer. 
} 
\endinsert 
\bigskip  
The focal instrumentation consists of two infrared recombiners. The 
first one uses a classical beamsplitter (see 4.1). The two recombined beams are 
focused on two pixels of a NICMOS III infrared camera (Millan-Gabet et al.,  
1999). This 
camera is able to read up to ten fringe frames per second. Each fringe frame, 
containing 256 samples, is made by scanning the optical path difference between 
the two beams with a mirror mounted on a 60-micron stroke piezo-electric 
transducer (PZT). Last scientific results obtained with this instrument include 
environment characterization of Herbig AeBe stars (Millan-Gabet et al., 1998) 
and dust shell diameter measurement of CI Cam (Traub et al., 1998). 
\bigskip 
The second recombiner named FLUOR (Fiber-Linked Unit for Optical 
Recombination, (Coud\'e du Foresto and Ridgway, 1992) consists of single-mode 
fiber optics interfering beams (Shaklan and Roddier, 1987). These fibers 
have been designed to propagate infrared light at K band (2.2 $\mu$m) in 
TEM mode only, like a coaxial cable. Therefore, only plane waves 
perpendicular to the axis of the fiber may propagate over long distances. 
Concretely, this results in a ``spatial filtering'', smoothing the wavefronts 
that have been corrugated by atmospheric turbulence. Figure 3 
depicts the principle of wavefront smoothing by spatial filtering with a 
pinhole and with a single-mode fiber. The advantage of such a technique for 
interferometry is a reduction of the uncertainty on the measured visibility. 
\bigskip 
\noindent 
\midinsert  
{\eightpoint  
\noindent 
\centerline{\psfig{figure=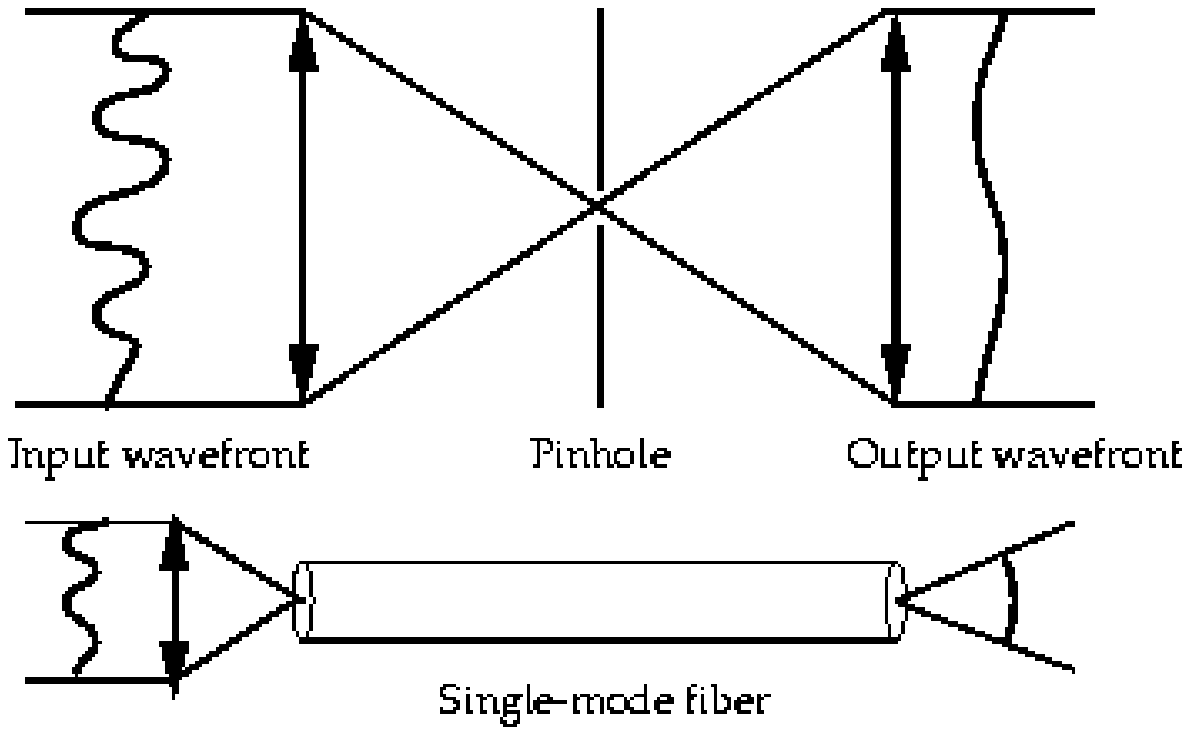,height=6cm,width=12cm}}  
\bigskip 
\noindent 
{\bf Figure 3.} Principle of wavefront smoothing by spatial filtering with a  
pinhole and with a single-mode fiber.  
} 
\endinsert 
\bigskip 
Drawbacks of spatial filtering are a loss of optical coupling efficiency and 
larger photometric variations due to the turbulence. The FLUOR experiment was 
originally set up at the McMath solar tower of the Kitt-Peak National 
Observatory (Arizona), with a 5~m baseline. It has been installed at IOTA since 
1994. The current FLUOR bench use a 180-micron stroke PZT for scanning the 
fringes. The detector used is the NICMOS III of IOTA. Four pixels are read (two 
for the interferometric fiber outputs, two for the photometric fiber outputs). 
Accurate diameter measurements of Mira variable stars (Perrin, 1999) and 
cepheids (Kervella et al., 1999), effective temperature measurements of giant 
stars (Perrin et al., 1998), have recently been done with FLUOR on IOTA. 
Figure 4 depicts the schematic of the FLUOR recombiner.  
\noindent  
\midinsert 
{\eightpoint   
\noindent 
\centerline{\psfig{figure=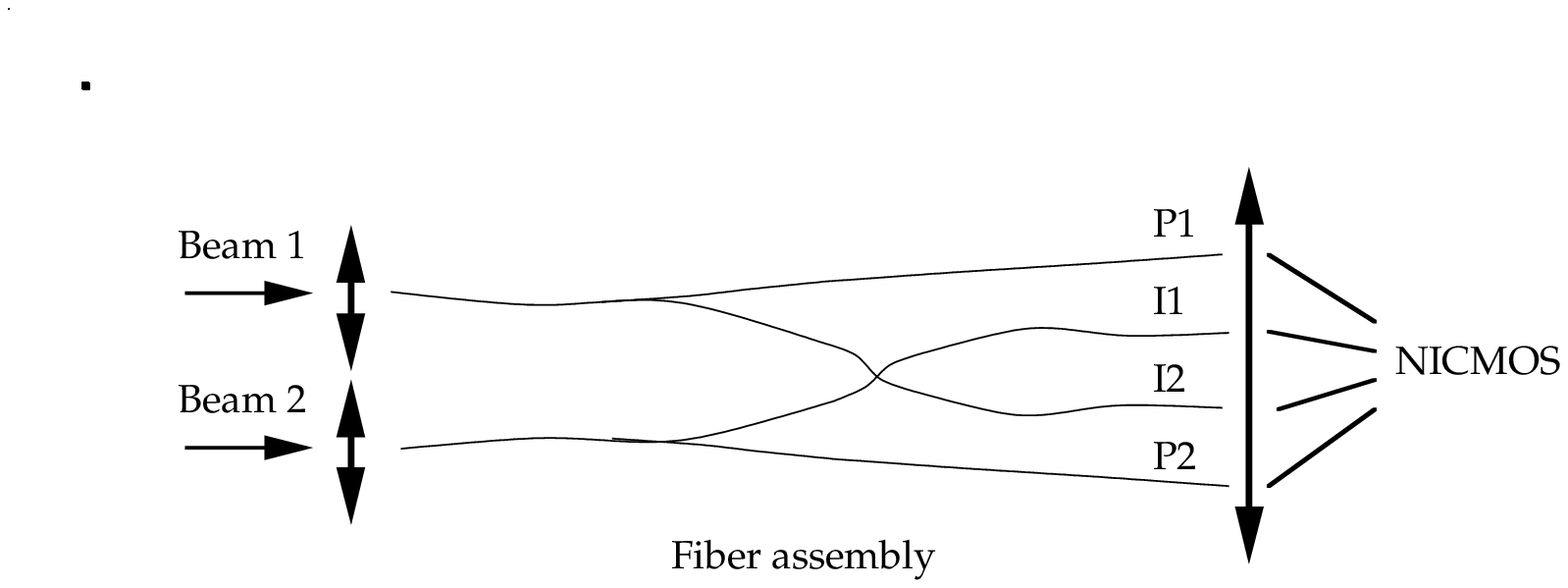,height=6cm,width=13cm}}  
\bigskip  
\noindent 
{\bf Figure 4.} 
Schematic of the FLUOR recombiner. P1 and P2 are the photometric output fibers. 
I1 and I2 are the interferometric output fibers. These outputs are imaged by a 
lens on a NICMOS infrared array detector.  
} 
\endinsert 
\bigskip  
Attempts to observe with single-mode fibers at longer wavelengths 
at IOTA (TISIS experiment) have been done in L band (3.5 $\mu$m) by Menesson et 
al., (1999), and in M band (5 $\mu$m) in March 1999 (Menesson et al., 2000). 
However, the thermal background of the instrument and the presence of an 
atmospheric ${\rm H}_2{\rm O}$ absorption line in M band barred  
fringe acquisition. 
\vskip 20 pt 
\noindent   {\bf 3.8. Optical very large array (OVLA)} 
\bigskip 
\noindent  
The OVLA is a project that was initiated about 10 years ago by A. Labeyrie. 
It is proposed to build an interferometer using innovative concepts. 
First, the collectors will be radically different from what has already been 
imagined. Each telescope structure (Dejonghe et al., 1998) is a 2.8 m diameter 
fiberglass sphere which may be oriented in any direction thanks to three motors 
featuring specially designed ``barrel-caster'' cabestans, mounted on their 
shafts. The sphere rests on these three barrel-casters: when one of them is 
steering the sphere, no friction occurs from the two other ones. The mirror of  
each telescope 
(1.5~m diameter with f/1.7) is made of ordinary window glass. It is 24~mm thick 
and weights 180 kg. An active optics is, therefore, required in order to get 
high-quality wavefronts for interferometric purpose: the mirror rests on three 
hard points and 29 actuators able to accurately correct its shape. 
Correction of the spherical aberration may be done by applying an electric 
current through the mirror coating between two chosen points of the edge, thus 
heating the top side of the mirror to compensate the noticed temperature 
difference with the bottom side (about 0.5$^\circ$C). A secondary mirror makes 
the beam afocal and compressed, a third steerable flat mirror sends this beam 
out through a slit located on the sphere. A motorized shutter can partially 
close this slit when a barrel-caster and the slit are in coincidence. It has  
been projected to mount the sphere and its motorization on a six-leg robot 
(hexapode) able to move on the ground while fringes are acquired (in order to 
compensate the OPD between beams). Due to its original features, an OVLA 
telescope requires a significant amount of electronics and a control system 
(Lardi\`ere et al., 1998) able to manage in real-time motors, actuators, 
shutter, hexapode. The first OVLA telescope has been tested in October 
1999 (Arnold et al., 2000): the 
point spread function of the optics was 4" FWHM. Improvements of the image 
quality by a better control of the actuators remains, therefore, to be done in 
order to obtain diffraction-limited images. Once this is done, the OVLA 
might be used as the third telescope of the GI2T interferometer 
becoming then the ``GI3T''. Figure 5 depicts the first operating telescope of 
proposed OVLA. 
\bigskip 
\noindent 
\midinsert 
{\eightpoint  
\noindent 
\centerline{\psfig{figure=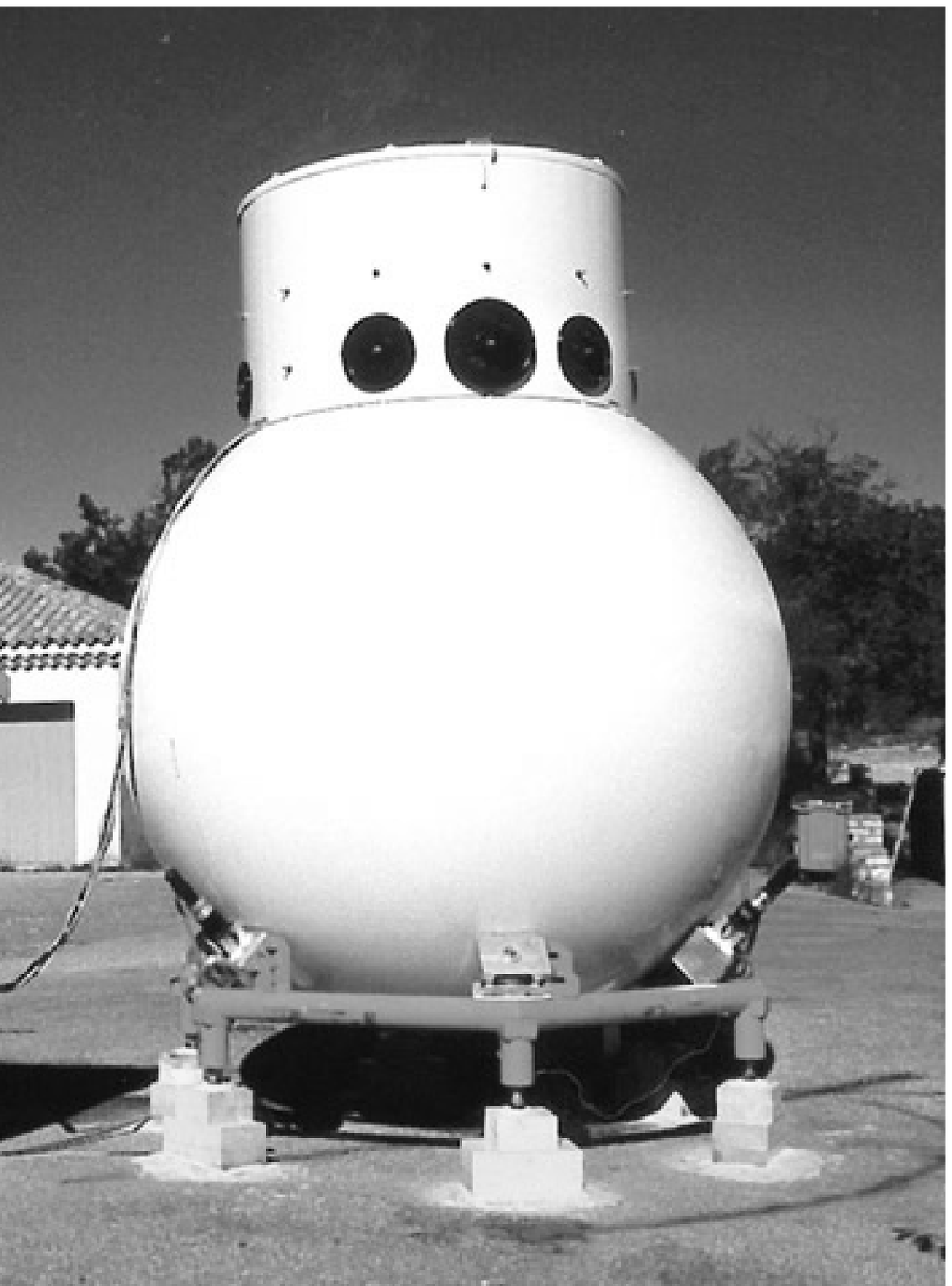,height=10cm,width=7.5cm}}  
\bigskip 
\noindent 
{\bf Figure 5.} One of the telescope of proposed OVLA (Courtesy: O. Lardi\`ere). 
} 
\endinsert 
\bigskip 
The OVLA has actually been thought for different possible aperture diameters 
including 12 to 25~m (Labeyrie, 1998c). A new telescope structure has been 
imagined for this class of very large collectors: the ``cage telescope'', in 
which the sphere is replaced with an icosahedral truss steerable by a different 
mechanical system. There are several 
options for the configuration of the OVLA interferometer featuring 27 apertures 
(like the Very Large Array radio-interferometer). The first one is to build a 
Fizeau interferometer, i.e. a fragmented giant telescope. Each telescope is 
shaped as it was a segment of the paraboloid mirror of the synthesized giant 
telescope. The array is arranged on the ground to form an ellipse. Images are 
then directly obtained at a recombination station located at a focus of the 
ellipse. This Fizeau configuration was thought for LOVLI (Lunar Optical Very 
Large Interferometer), the moon-based version of OVLA. However, controlling the 
particular shape of each telescope will be very difficult. A recent new concept 
of interferometric recombination, especially imagined for OVLA, is called 
``densified pupil'' (Labeyrie, 1996). In Fizeau mode, the ratio aperture 
diameter/separation  is constant from light collection to recombination in the 
image plane (homothetic pupil). In Michelson mode, this ratio is not constant 
since the collimated beams have the same diameter from the output of the 
telescope to the recombination lens. The distance between pupils is equal 
to the baseline at the collection and to a much smaller value just before the 
recombining lens. The disadvantage of the Michelson mode is a very narrow field 
of view compared to the Fizeau's. However, a densified pupil interferometer 
(``extreme Michelson mode''), where, in the recombination plane, the distance 
between two pupils corresponding to two telescopes is minimized to become about 
equal to their diameter, may be very interesting to get direct images without 
using the heavy procedure of aperture synthesis (visibility measurement, phase 
calibration, Fourier synthesis...). It can be demonstrated that the definition 
(i.e. number of pixels of the image) of a densified pupil interferometer is 
equal to the square of its number of apertures. One difficulty is cophasing all 
the beams. Since 27 ($=3^3$) telescopes are expected, the cophasing of the 
whole array may be done hierarchically (Pedretti and Labeyrie, 1999) by  
cophasing triplets of beams (yielding a honeycomb pattern in the image plane),  
then triplets of triplets, etc... The  limitation of the cophasing 
procedure by the photon noise would not be very important. According to 
numerical simulations, the expected limiting magnitude of the densified pupil 
OVLA is 8.3 if 10~cm apertures are used and 20 for 10~m apertures. 
\vskip 20 pt 
\noindent   {\bf 3.9. Mitaka optical infrared arrays (MIRA)} 
\bigskip 
\noindent  
The MIRA project (a collaboration between the University of Tokyo and the  
National Astronomical Observatory of Japan) does not consist of one but several 
interferometers built one-by-one, each instrument being an upgrade of the  
previous one. The first of the series was MIRA-I (Machida et al., 1998). It 
had 25~cm siderostats and a 4~m baseline. The fringe detector was designed for  
800~nm wavelength. Its successor, MIRA-I.2 (Sato et al., 1998) has the same  
baseline and slightly larger siderostats (30~cm). It features the equipment  
encountered on many operating interferometers: beam compressors (yielding 30~mm  
beams), delay-line operating in vacuum, tip-tilt correction system and laser  
metrology. MIRA-I and MIRA-I.2 are instruments specially designed for practicing 
interferometry and testing devices. The experience acquired from these  
interferometers will be useful for building larger interferometers of the MIRA  
project like MIRA-II, MIRA-SG and MIRA-III, which will be 
instruments for astrophysical research. 
\noindent 
\vskip 20 pt 
\centerline{\bf 4. Data acquisition and processing in optical interferometry} 
\bigskip 
\noindent   
Operating a long-baseline interferometer (i.e. finding fringes, 
measuring their visibility, interpreting the result) is a long and difficult 
process. First, a correct determination of the baseline vector must be 
established. Once this has been done, one knows, for a given object to observe, 
how to set the position of the optical delay-line to get fringes within, 
usually, a few hundred micron interval around the expected null-OPD point. 
Then, optics must be adjusted to avoid various aberrations and vignetting, 
which may be difficult to avoid when light is fed through long and narrow pipes. 
Then, fringes are searched by adjusting the delay-line position. However, once 
they are found, mechanical constraints on the instrument, errors on the pointing 
model, thermal drifts, various vibrations and atmospheric turbulence make the 
null-OPD point changing. The position of the delay-line (or any other delaying 
device in the optical path) must be adjusted in order to keep the fringes 
within the ``observation window'': usually, the error on the OPD must be less  
than the coherence length defined by: 
 
$$L_c={\bar\lambda^2\over\Delta\lambda}. \eqno(1)$$ 
 
\noindent   
Where $\bar\lambda$ is the mean wavelength observed and $\Delta\lambda$ is the  
spectral interval. This real-time control is called ``fringe-tracking''. 
\bigskip 
When enough fringe patterns have been recorded, the visibility 
may be extracted. Then, a set of measured visibilities obtained (i.e. samples  
in the Fourier plane $(u,v)$ corresponding to the image) allows to partially  
reconstruct the high-angular resolution image of the observed object. 
\vskip 20 pt 
\noindent   {\bf 4.1. Fringe acquisition and tracking} 
\bigskip 
\noindent    
For visible spectrum, three possible set-ups for fringe acquisition 
exist. In the first one (white fringes), the OPD is temporally modulated by a 
sawtooth  signal, using a fast and short-travel delaying device (usually, a 
reflector  mounted on a PZT). The intensity of the 
recombined beams describes, therefore, over the time a fringe pattern that is 
recorded by one or several mono-pixel detectors (photo-multipliers, avalanche 
photo-diodes, InSb photometers). Natural OPD drift due to the 
Earth-rotation can also be used for acquiring fringes, as it was done by the 
SOIRD\'ET\'E interferometer (Rabbia et al., 1990). The second method (channeled spectrum) 
consists of imaging the dispersed recombined beam on a linear detector (CCD or 
photon-counting camera). In the third one (dispersed fringes), beams are 
dispersed prior to be recombined. Unlike the two previous techniques, 
recombination is not done by overlapping the beams, but by focusing them with a 
common lens, like in the original Michelson stellar interferometer. The detector 
used is a 2-D photon-counting camera. In infrared, no photon-counting is 
possible with the current technology. It is, therefore, important to use as less 
pixels as possible in order to reduce the global readout noise. Hence, the 
``white'' fringes set-up will be preferably used for infrared observations. 
Figure 6 depicts the various possible set-ups for beam recombination and fringe 
acquisition. 
\bigskip 
\noindent 
\midinsert 
{\eightpoint   
\noindent 
\centerline{\psfig{figure=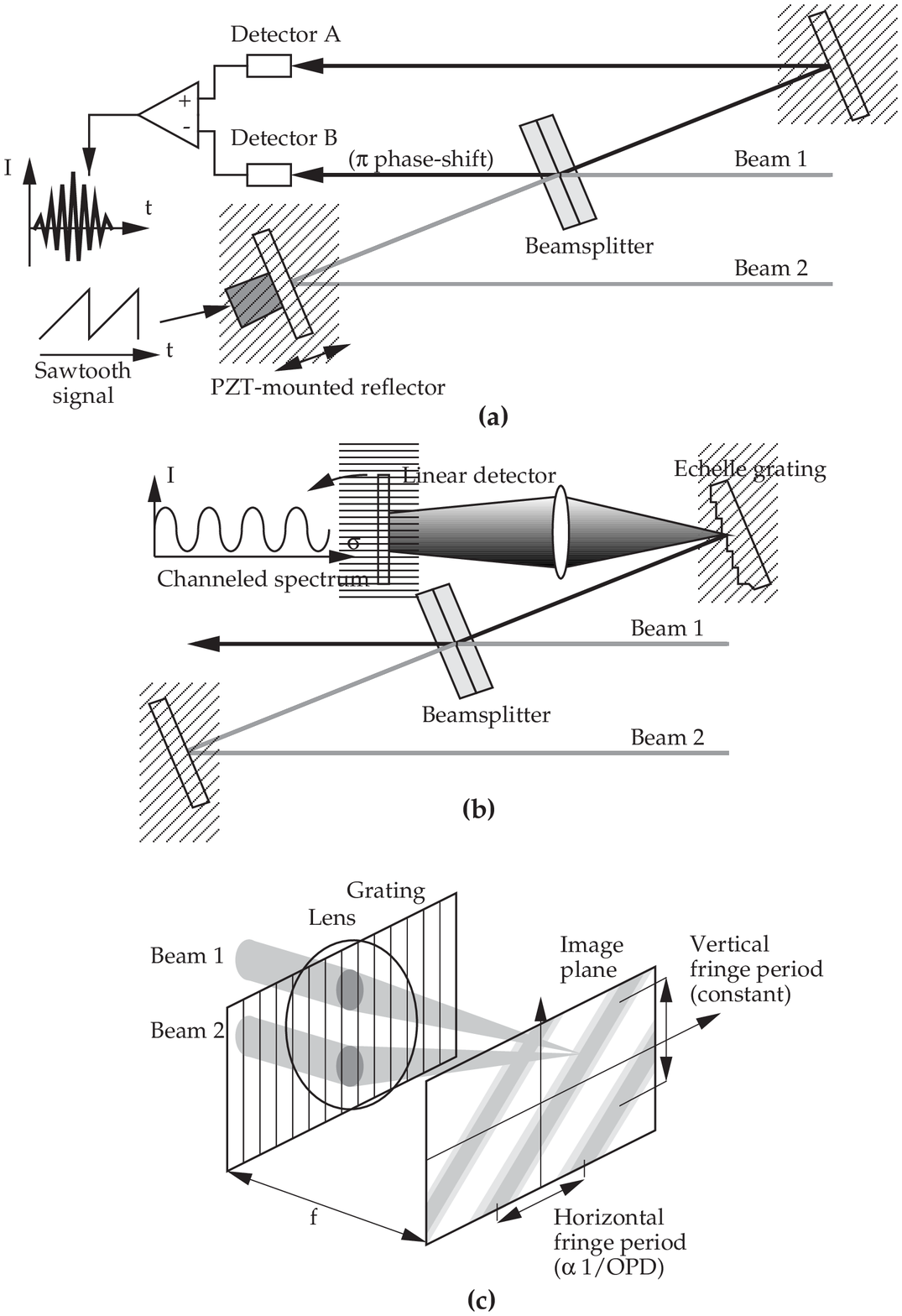,height=16.5cm,width=13cm}}  
\bigskip 
\noindent 
{\bf Figure 6.} 
Three possible set-up for beam recombination and fringe acquisition: white 
fringes (a), channeled spectrum (b), dispersed fringes (c).  
} 
\endinsert 
\bigskip  
Techniques to compensate the OPD drift between the two beams of a 
standard optical interferometer may be classified into two categories: 
coherencing and cophasing. The aim of coherencing is to keep the OPD within the 
coherence area of the fringes. Cophasing is a more demanding technique because 
the OPD must remain much smaller than the wavelength: fast compensation of the 
OPD variations due to the differential ``piston'' mode of the turbulence is, 
therefore, done, in order to ``freeze'' the fringes. 
\bigskip  
In white light set-up, the coherencing, as for IOTA (Morel et al., 2000), 
is done by scanning the OPD while acquiring interferometric signal, and then 
finding the null-OPD point in the fringe pattern. This yields the OPD correction 
to apply to the delay-line, at a few Hz servo-loop rate. With a channeled 
spectrum, the OPD  is proportional to the fringe frequency. The method used  
called ``group-delay  tracking'' (Lawson, 1995) is based on the Fourier  
transform of each frame acquired. The integration of the moduli of all the  
computed Fourier Transforms yields a peak whose position is proportional to the 
OPD. Group-delay 
tracking has been used on SUSI (Lawson, 1994) and COAST (Lawson, 1998) 
interferometers. A similar technique (Koechlin et al., 1996) named ``real-time 
active fringe-tracking'' (RAFT) has been applied to dispersed fringes on GI2T 
using a 2-D Fourier Transform. The advantage of RAFT over group-delay tracking 
is the knowledge of the sign of the OPD to measure and the possibility to be 
used with apertures larger than $r_0$, where overlapping wavefronts that have 
been corrugated by the turbulence would blur the fringes. However, dispersed 
fringes with large apertures require a complex optical system to rearrange the 
speckles in the image plane before dispersion and recombination (Bosc, 1988). 
Both group-delay tracking and RAFT allow a slow servo-loop period (up to a few 
seconds) by multiplying the coherence length by the number of spectral channels 
used. It is important to notice that their common use of the Fourier 
Transform make them optimal in the sense that they yield the same OPD than a 
maximum likelihood estimator (Morel and Koechlin, 1998). 
\bigskip  
Cophasing is usually performed with white fringes, using the 
``synchronous detection'' method, as it was used on the Mark III interferometer 
(Shao and Colavita, 1988): the OPD is quickly scanned over a wavelength. Signal 
acquired from the detector is then processed in order to yield the phase-shift  
to compensate and the visibility. This can be done easily (Shao and Staelin,  
1977) by integrating signal over four $\lambda/4$ bins, named $A$, $B$, $C$  
and $D$. Phase-shift and visibility modulus are then given by: 
 
$$\Delta\varphi=\arctan\left({B-D\over A-C}\right)\quad ;\quad 
V={\pi\sqrt{(A-C)^2+(B-D)^2}\over\sqrt{2}(A+B+C+D)}. \eqno(2)$$ 
 
The cophasing technique may be compared to adaptive optics, like coherencing 
with non-white fringes may be compared to active optics. We can notice that a 
compound method, based on the synchronous detection applied to signals from 
several spectral channels, has been used on the NPOI interferometer (Benson et 
al., 1998). Fringe-tracking is usually done from data acquired for scientific 
purpose (i.e. visibility extraction), in order to not ``share'' the photons 
between two instruments. Hence, the fringe signal-to-noise ratio (SNR) is  
optimal. This fringe SNR is given by the expression (Lawson, 1995): 
 
$${\rm SNR}\propto{NV^2\over\sqrt{1+0.5\times NV^2}}. \eqno(3)$$ 
 
\noindent  
Where $N$ is the number of photons acquired and $V$ is the visibility modulus. 
\bigskip  
However, it may be optimal to have two recombiners, one for visibility 
measurement, the other one for fringe-tracking. For example, at long baselines, 
when the expected fringe visibility is too low for tracking, it is possible to 
use a longer wavelength where the fringe contrast, for a white observed object, 
is higher. Meanwhile, fringes for computing the visibility are acquired at 
shorter wavelength than for tracking. Another method where photons are 
shared is called ``bootstrapping''. It consists in dividing the baseline into 
sub-baselines by adding apertures along. Fringe-tracking is performed on each 
sub-baseline, where the visibility is higher than with the whole baseline. 
Hence, fringes are tracked on the whole baseline as well. This method is used 
on the NPOI interferometer (Armstrong et al., 1998). Figure 7 depicts the 
principle of baseline bootstrapping. \noindent 
\midinsert 
{\eightpoint   
\noindent 
\centerline{\psfig{figure=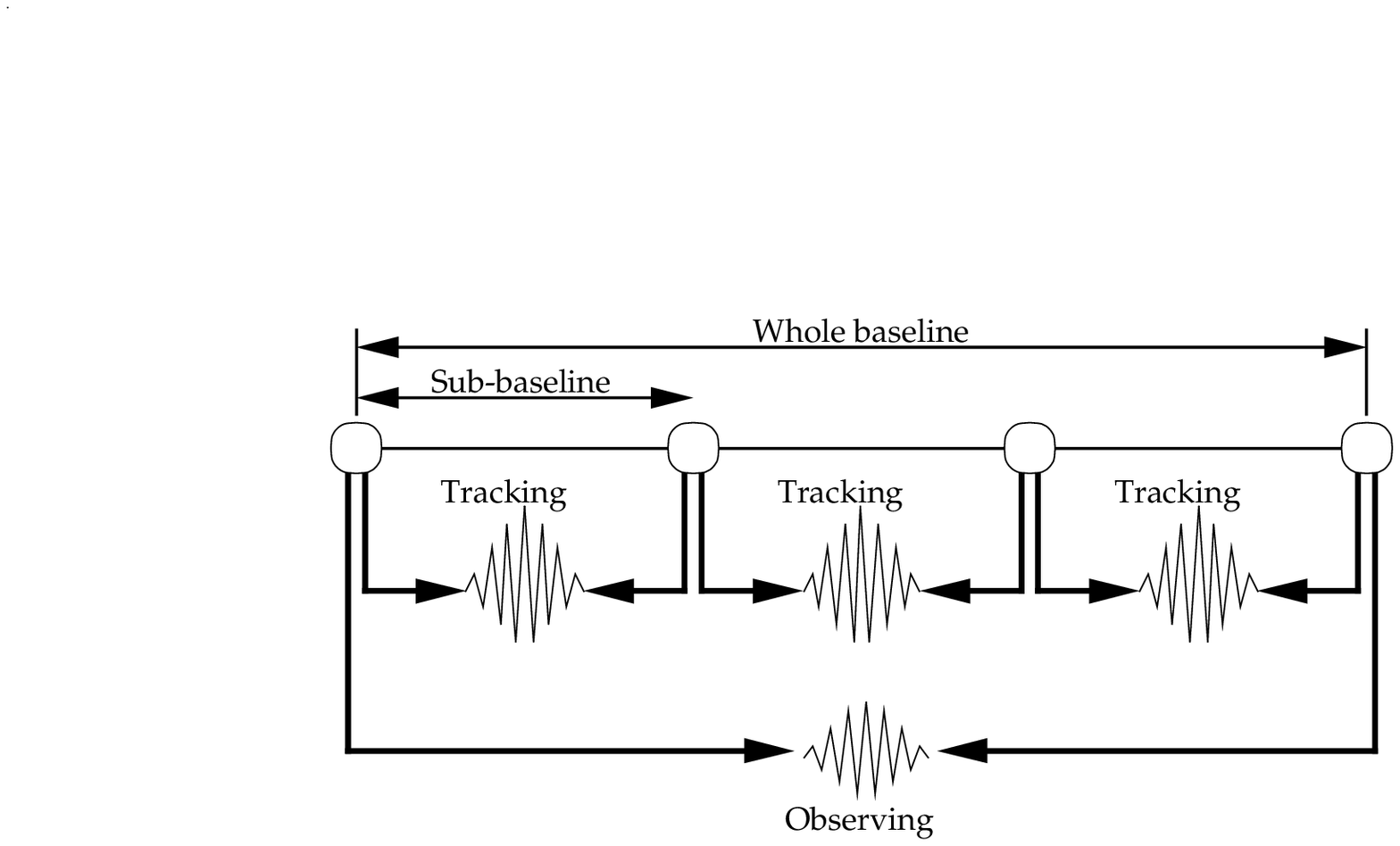,height=9cm,width=12cm}}  
\noindent 
{\bf Figure 7.} Principle of baseline bootstrapping. Apertures are represented  
by circles. 
} 
\endinsert 
Fringe-tracking methods may be enhanced by introduction of {\it a 
priori} information, in order to allow observations at fainter $V$ or fainter  
magnitudes. Gorham (1998) has proposed to improve white light cophasing by  
filtering data with a function computed to reduce the photon noise. 
The gain for the tracking limit magnitude, at constant $V$, is between 0.5 or 
0.7. Methods introducing {\it a priori} information for GDT or RAFT have been 
imagined as well (Padilla et al., 1998, Morel and Koechlin, 1998). 
\vskip 20 pt 
\noindent   {\bf 4.2. Data reduction} 
\bigskip 
\noindent    
The optimal integration time required for measuring a visibility 
point is a trade-off between the number of photons to collect and the Earth 
rotation shifting the sampled point in the $(u,v)$ plane. Most of the 
interferometers use two apertures and are unable to recover the complex 
visibility. Therefore, the information to extract from a batch of fringes is the 
modulus of the visibility. Theoretically, using merely the Fourier Transform 
would give an optimal estimate of the visibility modulus, as demonstrated by 
Walkup and Goodman (1973). However, white-light fringes obtained from  
coherencing are flawed by 
the differential piston that modulates their frequency. Techniques used in 
radio-interferometry (where wavelengths are much longer), like fitting a 
sinewave through the fringe data, are, therefore, not suitable. Perrin (1997)  
has proposed a method to remove the piston from fringes. However, this method 
requires a high fringe signal-to-noise ratio and may only be applied when fringe 
SNR is important. Schloerb et al., (1999) use a model of the turbulence effects 
to extract the visibility modulus. 
\bigskip  
Due to the atmospheric turbulence affecting the wavefronts before 
recombination, measurements of $V$ are biased by a random factor depending on 
the seeing quality. Instrumental flaws leading to optical aberrations and 
non-balanced flux between the two beams modify the measured visibility 
modulus as well. It is, therefore, important to calibrate each measure on an 
object by measuring $V$ on an non-variable unresolved source (e.g., a farther 
star) in the neighborhood of the studied object and at the same turbulence 
condition, i.e. right away after data acquisition on the studied object. To 
reproduce the instrumental conditions, the calibrator must roughly be as bright 
as the object to calibrate. 
\vskip 20 pt 
\centerline{\bf 5. Image reconstruction} 
\bigskip 
\noindent   
Interferometers with two apertures have limited possibilities for 
image reconstruction due to the absence of phase visibility recovering. Objects 
assumed with circular symmetry (``standard'' stars) may be reconstructed with 
two-aperture interferometers. However, a problem comes from the dark-limbening 
of the stars observed. The radial intensity profile of a star may be given 
(Hestroffer, 1997) by: 
 
$$I(r)=I(0)\left(1-{r^2\over R^2}\right)^{\alpha/2}. \eqno(4)$$ 
 
\noindent  
Where $R$ is the radius of the star and $\alpha$ is the dark-limbening 
factor depending on the stellar atmosphere. 
\bigskip  
Many interferometers cannot measure low visibilities existing at high angular  
frequency (i.e, when $\sqrt{u^2+v^2}$ is large), beyond the first zero of 
the visibility function. 
Reconstructions are, therefore, ambiguous and neither the diameter nor the 
dark-limbening factor may be accurately determined. Usually, $\alpha$ is an 
{\it a priori} information given by the stellar atmosphere model. The diameter  
is, therefore, deduced from $\alpha$ and the interferometric data. 
\bigskip 
Binary systems have already been characterized by optical stellar 
interferometry. The expression of two unresolved sources, i.e. a binary 
system, 
is $O({\bf x})=a\delta({\bf x}+{\bf x}_0)+b\delta({\bf x}+{\bf x}_0+{\bf 
x}_s)$, where $|{\bf x}_s|$ is the angular separation. The visibility modulus 
corresponding to this function at ${\bf u}=(u,v)$ is, therefore: 
 
$$|\widehat O({\bf u})|=\sqrt{(a-b)^2+4ab\cos^2(2\pi{\bf u}.{\bf x}_s)}.
\eqno(5)$$ 
 
It is then useful to use a technique called ``super-synthesis'': the $(u,v)$ 
plane is  swept during an observation lasting several hours, due to Earth 
rotation. If we note ${\rm B}'_{EW}$ and ${\rm B}'_{NS}$ the orthogonal  
East-West and North-South components of the baseline vector at the ground of an  
interferometer located at the terrestrial latitude $\theta_l$, the $(u,v)$  
point sampled from a star of 
declination $\delta_\ast$, when its hour angle is $H$, is given by: 
 
$$\left\{\eqalign{u=&({\rm B}'_{EW}\cos H-{\rm B}'_{NS}\sin\theta_l\sin H)/\lambda\cr 
v=&({\rm B}'_{EW}\sin\delta_\ast\sin H+{\rm B}'_{NS}\left(\sin\theta_l\sin\delta_\ast 
\cos H+\cos\theta_l\cos\delta_\ast\right))/\lambda\cr}\right. \eqno(6)$$ 
 
After a large variation of $H$, several visibility moduli are therefore measured 
at different $(u,v)$ points and allow to determine the parameters ($a$,$b$ and  
${\bf x}_s$) of the system by fitting the function described by Eq. 5. 
\bigskip 
Reconstruction of more complex images involves the knowledge of complex 
visibilities. The phase of a visibility may be deduced from closure-phase terms 
(Jennison, 1958) using three telescopes (i.e., wrapped sums of the phases of 
the three visibilities, including instrumental and atmospheric biases, sampled 
by the network at a given configuration). Closure-phase in optical  
interferometry has 
already been used (Baldwin et al., 1998, Hummel, 1998) for bright objects like 
the Capella binary system. Process after data acquisition consists of phase 
calibration and visibility phase reconstruction from closure-phase terms by 
techniques similar to bispectrum processing. From complex visibilities acquired 
from an array with a large number of aperture, it is possible to 
reconstruct the 
image by actually  interpolating the function in the $(u,v)$ plane. This has 
been done in  radio-interferometry for a few decades. The most popular 
algorithm for  reconstructing image from 
$(u,v)$ plane samples is named CLEAN (H\"ogbom, 1974). CLEAN subtracts 
iteratively, from the image given by the inverse Fourier Transform of the 
visibilities measured on the object (the ``dirty map''), a  fraction of the 
image 
given by the array from an unresolved source (the ``dirty beam'') centered on 
the maximum value of the dirty map. However, this simple  algorithm is subject 
to instability problems leading sometimes to wrong results. More robust versions 
of CLEAN have been designed (Cornwell, 1983, Dwarakanath et al., 1990). 
Among alternative algorithms for image reconstruction in aperture  synthesis are 
maximum entropy method (MEM) and WIPE (Lannes et al., 1997). It can be proved 
(Mar\'echal et al., 1997) that both MEM and WIPE derivate from a  common 
principle known as ``maximum entropy on the mean''. 
\vskip 20 pt 
\centerline{\bf 6. Space-borne interferometers} 
\bigskip 
\noindent  
As classical astronomy at visible and infrared wavelengths already did 
it (NASA's Hubble Space Telescope, ESA's Infrared Space Observatory), 
long-baseline optical and infrared interferometry will, in the next years, take 
advantage of observing from outer space (absence of atmospheric turbulence, 
observation possible at any wavelength and for long periods, easy cooling of 
optics and detectors). However, the first projects for a space interferometer 
(FLUTE, TRIO) were  proposed about two decades ago (Labeyrie et al., 1980, 
1982a, 1982b). The main difficulty was to develop a technology featuring 
high-precision 
positioning as well as toughness required  for space operation. Size and weight 
issues must also be addressed, depending on the chosen orbit. For example, if 
one wish to reach the Lagrangian point 2 (where the Sun and Earth gravitations 
are equal, stabilizing, therefore, any space-borne instrument), then the maximum 
payload mass of an Ariane V european launcher is 4998 kg. 
\vskip 20 pt 
\noindent  {\bf 6.1. Astrometry from space} 
\bigskip 
\noindent   
The accurate determination of star angular positions will provide 
crucial data for astrophysics. For example, precise parallax distances of 
cepheids will help to establish a period/absolute magnitude relationship in 
order to calibrate distances of galaxies, thus reducing the uncertainty on the 
value of $H_0$. For some galaxies, distance from Earth could be computed by 
tracking the stars of the lower and upper sides of the observed galaxies, 
yielding its apparent transverse velocity $v_t=v_0/D$, where $v_0$ is the 
actual edge velocity and $D$ the distance from Earth. Using spectroscopic 
ground-based measures giving the radial velocity 
$v_r=v_0.\sin i$ (where $i$ is the disk inclination), one can deduce $D$: 
 
$$D= {v_r\over v_t\sin i}. \eqno(7)$$ 
 
The quest for extra-solar planets (exo-planets) is another challenge for 
high-precision astrometry. A planet orbiting around a star causes a revolution 
of the star around the center of gravity defined by the two masses. Like galaxy 
velocity, this periodical short motion known as ``wobble'' has a radial 
counterpart measurable from ground by spectrometry. Thus, if Doppler-Fizeau 
effect measurements have already led to detect from Earth jovian planets around 
stars, like for example 51-Peg (Mayor and Queloz, 1995) or 47-Uma (Marcy 
and Butler, 1996), smaller planets might be detected by measuring the stellar 
photocenter motion due to the wobble. 
\bigskip  
Very valuable astrometry results from space have already been obtained 
by the Hipparcos satellite (Perryman, 1989). Hipparcos used phase shift 
measurement of the temporal evolution of the photometric level of two stars 
seen drifting through a grid. The successor of Hipparcos, Gaia (Lindengren and 
Perryman, 1996), will probably use the same technique with improvements, 
yielding 
more accurate results on a larger number of objects. However, only space-borne 
interferometers will achieve very high precision angular measurements. 
\bigskip For an interferometer consisting of two apertures separated by a 
baseline {\bf B}, the external optical delay $d$, while an object with altitude 
$\theta$ is observed in a broad spectral range (i.e. white light), is: 
 
$$d=|{\bf B}|\times \cos\theta. \eqno(8)$$ 
 
This delay can be deduced from the position of the optical delay-line of the 
instrument set up such that the central fringe of the interference pattern 
appears in a narrow observation window. The position, as well as $|{\bf B}|$, 
are measured by laser metrology. Hence, $\theta$ is deduced with a high 
precision. For a space-borne interferometer, the issue is to find a reference  
for the angle measured. Usually, a grid of far objects like quasars are used as 
a reference frame. Then, two modes of observation are possible: the  
``wide-angle'' 
and the ``narrow-angle'' modes. In wide-angle mode, the large angle difference 
between the reference and the studied object usually requires collector motions. 
In narrow-angle, the two objects are in the field of view of the instrument, 
therefore, no motions are required and the accuracy of the measurement is 
improved. However, it is difficult to have always a correct reference star 
within the field of view for any studied object. Narrow-angle astrometry is, 
therefore, more suitable for wobble characterization. Figure 8 depicts the 
principle of an interferometer for astrometry. 
\bigskip 
\noindent 
\midinsert 
{\eightpoint   
\noindent 
\centerline{\psfig{figure=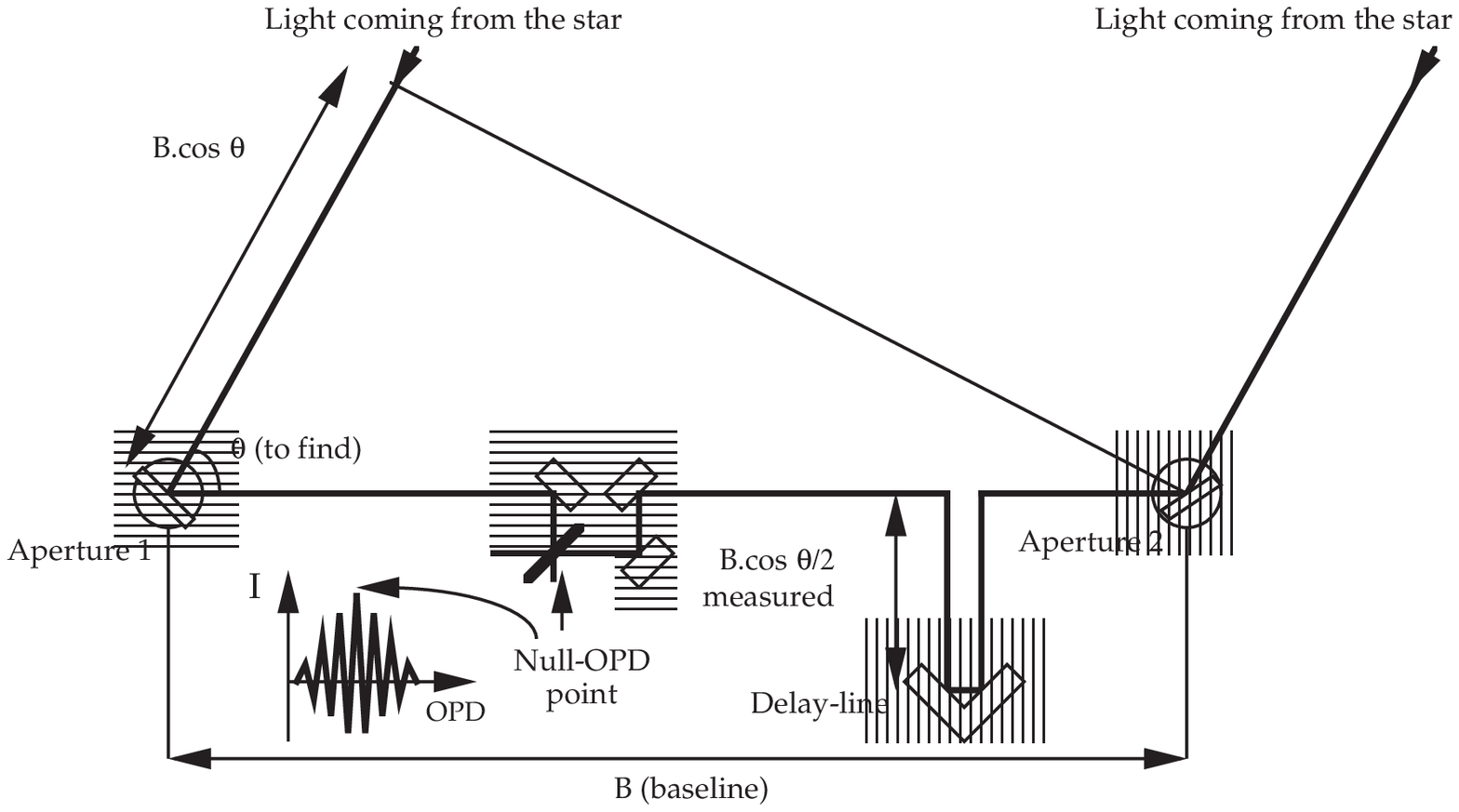,height=7cm,width=12.5cm}} 
\bigskip 
\noindent 
{\bf Figure 8.} Principle of an interferometer for astrometry.   
} 
\endinsert 
\vskip 20pt 
\noindent  {\bf 6.2. First space-borne interferometers} 
\bigskip 
\noindent   
The expected pioneer of this new generation of scientific spacecrafts 
is ``Space Technology 3'', or ST3 (Gorham et al., 1999), formerly known as 
``Deep Space 3''. This NASA's mission, scheduled for 2003, consists of two 
independent free-flying elements launched into an Earth-trailing heliocentric 
orbit. One is a collector sending light from the observed object to the second 
element featuring another collector, an optical delay-line and a beam 
recombiner. The aim of ST3 is the demonstration and validation of technologies 
that might be used for future space-borne interferometers like SIM or TPF (see 
further). Thus, the two elements of ST3 should be able to move up to 1 km from 
each other, thanks to ionic micro-engines, while being controlled by a laser 
metrology. However, the designed delay-line of ST3 can delay up to 20~m of 
optical pathlength only. A 200~m maximum projected baseline would, therefore, be 
possible with 1~km spacecraft separation. More than just a technology 
experiment, ST3 will be used as an imaging interferometer for studying 
Wolf-Rayet or Be stars (Linfield, 1999). 
\bigskip  
After ST3, SIM (Space Interferometry Mission) will be the next 
space-borne interferometer built and launched by NASA (Unwin et al., 1998). The 
main goal of SIM will be a collect of new high-precision astrometry results 
(see above), including the possibility of jovian planet detection around stars  
up to 
1 kilo-parsec distant and terrestrial planet detection around nearby stars. The 
final design of SIM, known as ``SIM Classic'' has recently been decided (Unwin, 
1999). It consists of one free-flyer with a 10~m boom supporting 30~cm 
collectors. The expected angular accuracy is 1~$\mu$as in narrow-angle mode 
(with a 1$^\circ$ field of view) and 4 $\mu$as in wide-angle mode. The 
sensitivity for astrometry is $m_V=20$ after four hour integration. SIM will 
work in the visible spectrum (0.4 to 0.9~$\mu$m). In order to get an accurate 
knowledge of the baseline vector {\bf B} for wide-angle astrometry without 
collector motions, SIM will feature two auxiliary interferometers, aimed at 
reference stars (``grid-locking''). The schematic design of the SIM is depicted 
in figure 9.  
\bigskip  
\noindent  
\midinsert 
{\eightpoint   
\noindent 
\centerline{\psfig{figure=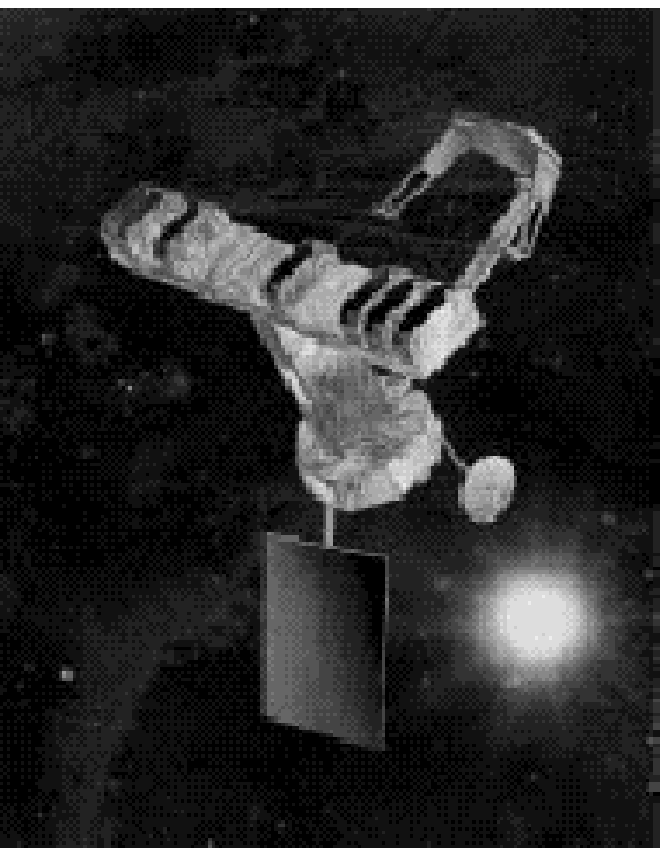,height=8cm,width=7cm}}  
\bigskip 
\noindent 
{\bf Figure 9.} The schematic design of the SIM (Courtesy: NASA/JPL/Caltech). 
} 
\endinsert 
\bigskip  
Besides its abilities for astrometry, SIM will feature a nulling mode. 
Like coronography, the nulling is a technique used for masking a bright central 
object in order to reveal its fainter environment (Bracewell and Macphie, 1979). 
Basically, a nulling system is an interferometer with a $\pi$ phase shift 
introduced in one beam. Therefore, the central fringe of the interference  
pattern is dark, allowing the fringe pattern from a faint object to 
appear. The quality of a nulling is defined by the ``null depth'' $N$: 
 
$$N=(1-V\cos\varphi_e)/2\approx(\pi\sigma_{\rm OPD}/\lambda)^2. \eqno(9)$$ 
 
\noindent  
Where $V$ is the fringe visibility modulus, $\varphi_e$ the phase error between 
the two recombined beams and $\sigma_{\rm OPD}$ the standard deviation of the 
optical path difference between the two beams. Nulling at $N=25,000$ has been 
obtained with laser light in laboratory (Serabyn, 1999). The nulling system of 
SIM is expected to reach $N=10,000$ with white light. 
\bigskip  
SIM should be launched in 2005 into an Earth-trailing heliocentric 
orbit and be operational from 2006 for a five-year mission. 
\vskip 20 pt 
\noindent  {\bf 6.3. Searching for life on other planets} 
\bigskip 
\noindent   
The old, somehow philosophical, question ``Is Earth the only planet 
sheltering life or not?'', might be answered in the next decades, thanks to 
interferometry. The knowledge of the chemical composition of any planetary 
atmosphere gives hints about the likeliness to find carbon-based life, as we 
know it, on this planet. Lovelock (1965) has suggested that the simultaneous 
presence on Earth of a highly oxidized gas, like ${\rm O}_2$, and a highly 
reduced gas, like ${\rm CH}_4$ and 
${\rm N}_2{\rm O}$ is the result of the biochemical activity. However, finding 
spectral signatures of these gases on an exo-planet would be very difficult. An 
alternative life indicator would be ozone (${\rm O}_3$), detectable as an 
absorption line at 9.6~$\mu$m. On Earth, ozone is photochemically produced from 
${\rm O}_2$ and, as a component of the stratosphere, is not masked by other 
gases. Finding ozone would, therefore, indicate a significant quantity of ${\rm 
O}_2$ that should have likely been  produced by photosynthesis (L\'eger et al., 
1993). Moreover, for a star like the Sun, detecting ozone can be done 1000 times 
faster than detecting ${\rm O}_2$ at 0.76 $\mu$m: estimates made by Angel and 
Woolf (1997) show that the requirements for planet detection in the visible 
with an 8~m telescope are not matchable with current technology. 
\bigskip  
One of the imagined instruments for ozone search on exo-planets is 
``Darwin'' (Penny et al., 1998), a.k.a. IRSI (InfraRed Space 
Interferometer), a project selected by ESA as a ``cornerstone mission''. The 
aim is the discovery and characterization of terrestrial planet systems around 
nearby stars (closer than 15~pc) by direct detection (i.e. involving the 
detection of photons from the planet and not from the star as it is done with 
Doppler-Fizeau effect detection or wobble detection). The design of this 
instrument has not been established yet, but some features will likely be found 
in the final version of Darwin. 
\bigskip  
Basically, Darwin will have to overcome two major 
difficulties for achieving Earth-like planet detection. The first one concerns 
stellar light quenching. Interferometric nulling techniques will obviously be 
employed to address this issue. Severe requirements about the optical 
quality of the nulling device might involve spatial filtering (Ollivier 
and Mariotti, 1997), by pinhole or single-mode fibers, to smooth the beam 
wavefronts. The second difficulty is the expected presence of exo-zodiacal light 
(infrared emission from the dust surrounding the observed star). 
\bigskip  
Several solutions for the instrument design have, therefore, been imagined. The 
first one consists of five free-flying collectors (1 to 2~m telescopes) 
rotating around a central recombiner. The image of the exo-planetary system is  
constructed after a 2$\pi$ rotation of the system. The odd number of apertures  
enables a recovery of 
the signal from a planet (which is an asymmetrical object from the axis of 
rotation defined by the star) drowned in the signal from the exo-zodiacal disk 
(which is a symmetrical object around the star). An alternative configuration, 
recently imagined, consists of six collectors arranged to form a triangle 
(Mariotti and Menesson, 1998). In this configuration, no rotation of the 
system is 
required. Darwin is expected to be launched in 2009 into an orbit at 5~AU from 
the sun, in order to reduce the illumination by solar zodiacal light. 
\bigskip  
A project named TPF (Terrestrial Planet Finder), very similar to Darwin/IRSI is 
currently studied by NASA (Beichman, 1998). Like Darwin, the final 
design has not been decided yet. The current version (Lawson et al., 1999) 
features four aligned 3.5 m free-flying telescopes and a central recombiner. The 
baseline from the two most separated telescopes can span from 75~m to 1~km. Like 
the original design of Darwin, the collectors of TPF will rotate around the 
recombiner for planet detection. In this case the maximum baseline is 135~m. For 
planet imaging, telescopes will move along parallel straight lines and could be 
separated by 1~km. Instrumentation for spectroscopy on TPF will include a 
$R=30$ spectrometer working between 7~$\mu$m and 20~$\mu$m for planet detection, 
and a $R=300$ spectrometer working between 3~$\mu$m and 30~$\mu$m for imaging. 
The expected time to find a planet and then to determine whether ozone is 
present in its atmosphere should be 15 days for each star zeroed in on. The  
launch of TPF is expected in 2010. Its five-year mission will start one year  
later. Instead of a far orbit location as Darwin is supposed to reach, TPF  
will be placed on a 
Earth-trailing orbit or at the Lagrangian point 2. Despite the problem of solar 
zodiacal light (that is expected to be overcome by using telescopes larger than 
Darwin's), such orbits provide easier radio-transmissions, a larger available 
solar power and a heavier payload possible for the launcher. 
Figure 10 depicts the conceptual design of TPF 
\bigskip 
\noindent 
\midinsert 
{\eightpoint   
\noindent 
\centerline{\psfig{figure=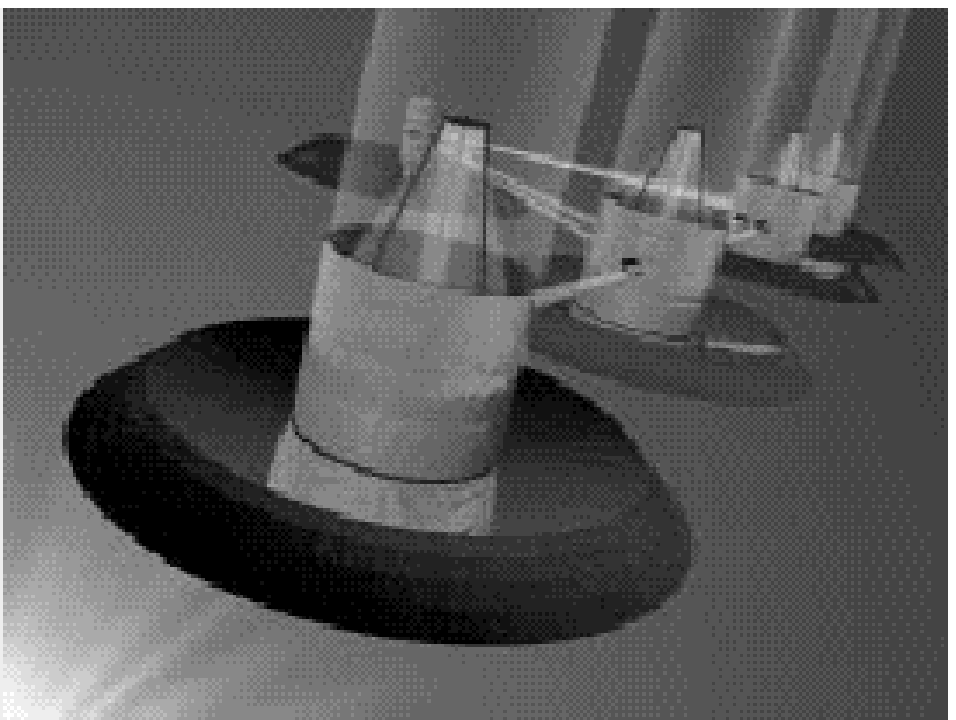,height=10cm,width=9cm}}  
\bigskip 
\noindent 
{\bf Figure 10.} The conceptual design of the TPF (Courtesy: NASA/JPL/Caltech). 
} 
\endinsert 
\vskip 20pt 
\noindent {\bf 6.4. Long-term perspective} 
\bigskip 
\noindent Space-borne interferometry projects for years spanning from 2020 to 
2050 already exist. However, the reader should be aware that such projects must 
be regarded as drafts for future instruments. No one can forecast, today, how 
future space-borne interferometers will actually look like. 
\bigskip  
For the post-TPF era, NASA has imagined an enhanced version featuring 
four 25~m telescopes and a $R\geq 1000$ spectrometer. This interferometer would 
be able to detect on an exo-planet lines of gases directly produced by 
biochemical activity. The next step proposed by NASA is an array of 25 
telescopes, 40~m diameter each, that would yield 25 
$\times$ 25 pixel images of an Earth-like planet at 10 pc, revealing its 
geography and eventually oceans or chlorophyll zones. 
\bigskip  
A comparable project has been proposed by Labeyrie (1999). It consists 
of 150 telescopes, 3~m diameter each, forming an interferometer with a 150~km 
maximum baseline. Such an instrument would give a 40 $\times$ 40 pixel image of 
an Earth-like planet at 3~pc, providing the same information as described 
previously. 
\vskip 20 pt 
\centerline{\bf 8. Conclusions} 
\bigskip 
\noindent  
The angular resolution of any stellar object in the visible wavelength can  
vastly be improved by using long baseline interferometry.  
The angular diameter for more than 50 stars have been measured 
(DiBenedetto and Rabbia, 1987, Mozurkewich et al., 1991, Dyck et al., 1993) with 
accuracy better than 1\% in some cases with the ground-based long baseline 
amplitude interferometers at optical and IR wavelengths. 
Apart from the measurements of the diameters, distances, masses 
on the stellar surfaces, among others, this technique can detect the 
morphological details, viz., (i) spots and flares, (ii) granulations,  
(ii) oblateness etc., of giant stars. Eclipsing binaries (Algol type) which  
show evidence of detached gas rings  
around the primary are also good candidates for long baseline interferometry. 
The potential of this type of interferometer can be envisaged in determining the 
fundamental astrophysical informations of circumstellar envelopes such as, the 
diameter of inner envelope, colour, symmetry, radial profile etc. The objectives 
of very large array in optical/IR wave bands range from detecting other  
planetary systems to imaging the black hole driven central engines of quasars 
and active galaxies. 
\bigskip  
Several long baseline interferometers are either in operation or under 
development at various stages. Rapid increase in the scientific output at  
optical, as well as at infrared wave bands using these interferometers can  
be foreseen at the begining of this millennium. With improved technology, the 
long baseline interferometric arrays of large telescopes fitted with 
high level adaptive optics system that applies dark speckle coronograph 
(Boccaletti et al., 1998) may provide snap-shot images at their 
recombined focus using the concept of densified-pupil imaging (Pedretti and 
Labeyrie, 1999), and yield improved images and spectra of objects. One of the 
key areas where the new technology would make significant contributions is the 
astrometric detection and characterization of exo-planets. 
\bigskip   
However, the role of 
smaller interferometers should not be neglected, since such instruments could 
be useful for long observations of binary systems, testing new focal  
instrumentation designed for larger interferometers, or observing at short  
wavelengths (blue, UV). Moreover, this class of interferometers would be easily 
accessible to a broader community 
of astronomers and might be employed as education tools.   
\bigskip   
On the other hand, space-borne LBIs would provide the best ever spatial 
resolution of faint objects as the fringes can be recorded with longer  
integration time. The greatest advantage of such a project is being the  
absence of atmospheric turbulence. The bright prospects of LBI programmes can 
be witnessed from the future space interferometers like SIM, ST3 and Darwin 
which are effectively funded projects that will become an essential tool at the 
cutting edge of astronomical research in the new millennium. 
\vskip 20pt 
\noindent {\bf Acknowledgments}: 
The authors thank Dr P. R. Lawson at Jet Propulsion Laboratory, USA and Dr O.  
Lardi\`ere at Observatoire de Haute Provence, France, for photographs used 
in this paper. The service rendered by Mr. B. A. Varghese, Indian Institute of 
Astrophysics, Bangalore, India is also gratefully acknowledged. One of the  
authors Dr S. Morel is grateful to DGA-DRET (the scientific research office 
of the French Ministry of Defense) for having funded his post-doctoral stay  
at IOTA, and to JPL for its useful documentation and lectures. 
\vskip 20pt 
\centerline{\bf References} 
\bigskip {\eightpoint\parindent=0pt\everypar={\hangindent=0.5 cm} 
Anderson J. A., 1920, Ap. J., 51, 263. 
 
Angel R., Woolf N. J., 1997, Ap J, 475, 373. 
 
Armstrong J. T., Hummel C. A., Mozurkewich D., 1992a, Proc. ESO-NOAO conf. 
`High Resolution Imaging Interferometry', eds., J. M. Beckers \& F. Merkle,  
Garching bei M\"unchen, Germany, 673. 
 
Armstrong J. T., Mozurkewich D., Pauls T. A., Hajian A. R., 1998, Proc.  SPIE., 
conf. `Astronomical Interferometry', 3350, 461. 
 
Armstrong J. T., Mozurkewich D., Rickard L. J., Hutter D. J., Benson J. A., 
Bowers P. F., Elias N. M., Hummel C. A., Johnston K. J., Buscher D. F., Clark  
J. H., Ha L., Ling L. -C., White N. M., Simon R. S., 1998,  Ap J, 496, 550. 
 
Armstrong J. T., Mozurkewich D., Vivekanand M., Simon R. S., Denison C. S., 
Johnston K. J., Pan X. -P., Shao M., Colavita M. M., 1992b, A J, 104, 241. 
 
Arnold L., Labeyrie A., Mourard D., 1996, Adv. Space Res., 18, 1149. 
 
Arnold L., Lardi\`ere O., Dejonghe J., 
2000, Proc. SPIE, conf. `Interferometry in Optical Astronomy', 4006, (in preparation). 
 
Ayers G. R., Dainty J. C., 1988, Opt. Lett., 13, 457. 
 
Babcock H. W., 1953, PASP, 65, 229. 
 
Baldwin J. E., Beckett R. C., Boysen R. C., Burns D., Buscher D. F., Cox G. C., 
Haniff C. A., Mackay C. D., Nightingale N. S., Rogers J., Scheuer P. A. G., 
Scott T. R., Tuthill P. G., Warner P. J., Wilson D. M. A., Wilson R. W., 
1996, A \& A, 306, L13. 
 
Baldwin J. E., Boysen R. C., Haniff C. A., Lawson P. R., Mackay C. D., Rogers 
J., St-Jacques D., Warner P. J., Wilson D. M. A., Young J. S., 1998, Proc. 
SPIE., conf. `Astronomical Interferometry', 3350, 736. 
 
Baldwin J. E., Haniff C. A., Mackay C. D., Warner P. J., 1986, Nature, 320, 
595. 
 
Bedding T. R., 1999, astro-ph/9901225, PASP (to appear). 
 
Bedding T. R., Robertson J. G., Marson R. G., 1994, A \& A, 290, 340. 
 
Bedding T. R., Robertson J. G., Marson R. G., Gillingham P. R., Frater R. H., 
O'Sullivan J. D., 1992, Proc. ESO-NOAO, conf. `High Resolution Imaging 
Interferometry', eds. J. M. Beckers \& F. Merkle, Garching bei M\"unchen,  
Germany, 391. 
 
Beichman C. A., 1998, Proc. SPIE, conf. `Astronomical Interferometry', 3350,  
719. 
 
Benson J. A., Mozurkewich D., Jefferies S. M., 1998, Proc. SPIE, conf. 
`Astronomical Interferometry', 3350, 493. 
 
Bester M., Danchi W. C., Degiacomi C. G., Townes C. H., 1991, Ap J, 367, L27. 
 
Boccaletti A., Moutou C., Labeyrie A., Kohler D., Vakili F., 1998, A \& A, 
340, 629. 
 
Boden A. F., Koresko C. D., Van Belle G. T., Colavita M. M., Dumont P. J., 
Gubler J., Kulkarni S. R., Lane B. F., Mobley D., Shao M., Wallace J. K., 
1999, Ap J, 515, 356. 
 
Bosc I., 1988, Proc. ESO-NOAO, conf. `High-Resolution Imaging by 
Interferometry', ed. F. Merkle, Garching bei M\"unchen, Germany, 735. 
 
Bracewell R. N., Macphie R. H., 1997, Icarus, 38, 136. 
 
Brown R. H., 1974, `The Intensity Interferometry, its Applications to 
Astronomy', Taylor \& Francis, London. 
 
Brown R. H. and Twiss R. Q., 1958, Proc. Roy. Soc. A., 248, 222. 
 
Brown R. H., Davis J. and Allen L. R., 1967, MNRAS, 137, 375. 
 
Brown R. H., Jennison R. C. and Das Gupta M. K., 1952, Nature, 170, 1061. 
 
Burns D., Baldwin J. E., Boysen R. C., Haniff C. A., Lawson P. R., Mackay 
C. D., Rogers J., Scott T. R., Warner P. J., Wilson D. M. A., Young J. S.,  
1997, MNRAS, 290, L11. 
 
Burns D., Baldwin J. E., Boysen R. C., Haniff C. A., Lawson P. R., Mackay 
C. D., Rogers J., Scott T. R., St-Jacques D., Warner P. J., Wilson D. M. A.,  
Young J. S., 1998, MNRAS, 297, 467. 
 
Busher D. F., Haniff C. A., Baldwin J. E., Warner P. J., 1990, MNRAS., 245, 7. 
 
Butler P. R., Marcy G. W., 1996, Ap J, 464, L153. 
 
Colavita M. M., Boden A. F., Crawford S. L., Meinel A. B., Shao M.,  Swanson P. 
N., Van Belle G. T., Vasist G., Walker J. M., Wallace J. P., Wizinowich P. L., 
1998, Proc. SPIE, conf. `Astronomical Interferometry',  3350, 776. 
 
Colavita M. M., Wallace J. K., Hines B. E., Gursel Y., Malbet F., Palmer D. L., 
Pan X. P., Shao M., Yu J. W., Boden A. F., Dumont P. J., Gubler J., Koresko C. 
D., Kulkarni S. R., Lane B. F., Mobley D. W., Van Belle G. T., 1999, A J, 510, 
505. 
 
Cornwell T. J., 1983, A \& A, 121, 281. 
 
Coud\'e du Foresto V., Ridgway S.T., 1992, Proc. ESO-NOAO, conf. `High 
Resolution Imaging Interferometry', eds. J. M. Beckers and F. Merkle, Garching  
bei M\"unchen, Germany, 731. 
 
Davis J., Tango W. J., Booth A. J., Minard R. A., Brummelaar t. T. A., 
Shobbrook R. R., 1992, Proc. ESO-NOAO, conf. `High Resolution Imaging  
Interferometry', eds. J. M. Beckers and F. Merkle, Garching bei M\"unchen, Germany, 741. 
 
Davis J., Tango W. J., Booth A. J., O'Byrne J. W., 1998, Proc. SPIE, conf. 
`Astronomical Interferometry', 3350, 726. 
 
Davis J., Tango W. J., Booth A. J., Thorvaldson E. D., Giovannis J., 1999, 
MNRAS, 303, 783. 
 
Dejonghe J., Arnold L., Lardi\`ere O., Berger J.-P., Cazal\'e C., 
Dutertre S., Kohler D., Vernet D., 1998, Proc. SPIE, conf. `Advanced Technology 
Optical/IR Telescopes', 3352, 603. 
 
Derie F., Ferrari M., Brunetto E., Duchateau M., Amestica R., Aniol P., 
2000, Proc. SPIE, conf. `Interferometry in Optical Astronomy', 4006, (in preparation). 
 
DiBenedetto G. P.,  Conti G., 1983, Ap J, 268, 309. 
 
DiBenedetto G. P., Rabbia Y., 1987, A \& A, 188, 114. 
 
Dwarakanath K. S., Deshpande, A. A., Udaya Shankar N., 1990, J. Astr. Astron, 
11, 311. 
 
Dyck H. M., Benson J. A., Carleton N. P., Coldwell C. M., Lacasse M. G., 
Nisenson P., Panasyuk A. V., Papaliolios C. D., Pearlman M. R., Reasenberg R. 
D., Traub W. A., Xu X., Predmore R., Schloerb F. P., Gibson D., 1995, A J, 109, 
378. 
 
Dyck H. M., Benson J. A., Ridgway S. T., 1993, PASP, 105, 610. 
 
Dyck H. M., Benson J. A., Van Belle G. T., Ridgway S. T., 1996b, A J, 111, 
1705. 
 
Dyck H. M., Van Belle G. T., Benson J. A., 1996a, A J, 112, 294. 
 
Dyck H. M., Van Belle G. T., Thomson R. R., 1998, A J, (to appear). 
 
Falcke H., Davidson K., Hofmann K. -H., Weigelt G., 1996, A \& A, 306, L17. 
 
Faucherre M., Bonneau D., Koechlin L., Vakili F., 1983, A \& A, 120, 263. 
 
Fizeau H., 1868, C. R. Acad. Sci. Paris, 66, 934. 
 
Fried D. C., 1966, J. Opt. Soc. Am., 56, 1972. 
 
Goodman J. W., 1968, Introduction to Fourier optics, McGraw Hill Book Co., New-York. 
 
Gorham P. W., 1998, Proc. SPIE, conf. `Astronomical Interferometry', 3350, 116. 
 
Gorham P. W., Folkner W. M., Blackwood G. H., 1999, conf. `Working on the  
fringe', Dana Point, USA, to be published in ASP Conference Series, eds. S.  
Unwin and R. Stachnik. 
 
Grieger F., Weigelt G., 1992, Proc. ESO-NOAO, conf. `High Resolution Imaging 
Interferometry', eds. J. M. Beckers and F. Merkle, Garching bei M\"unchen,  
Germany, 481. 
 
Hajian A. R., Armstrong J. T., Hummel C. A., Benson J. A., Mozurkevich D., 
Pauls T. A., Hutter D. J., Elias N. M., Johnston K. J., Rickard L. J., 
White  N. M., 1998, Ap J, 496, 484. 
 
Haniff C. A., Busher D. F., Christou J. C., Ridgway S. T., 1989, MNRAS, 
241, 694. 
 
Haniff C. A., Mackay C. D., Titterington D. J., Sivia D., Baldwin J. E., Warner 
P. J., 1987, Nature, 328, 694. 
 
Harmanec P., Morand F., Bonneau D., Jiang Y., Yang S., Guinan E. P., Hall 
D. S., Mourard D., Hadrava P., Bozic H., Sterken C., Tallon-Bosc I., Walker G.  
A. B., McCook P. M., Vakili F., Stee P., 1996, A \& A, 312, 879. 
 
Hestroffer D., 1997, A \& A, 327, 199. 
 
The Hipparcos catalogue, 1997, ESA, SP-1200. 
 
H\"ogbom J. A., 1974, A \& AS, 15, 417. 
 
Hummel C. A., 1994, IAU Symp. 158, `Very high resolution imaging' ed., J. G. 
Robertson and W. J. Tango, 448. 
 
Hummel C. A., 1998, Proc. SPIE, conf. `Astronomical Interferometry', 3350, 483. 
 
Hummel C. A., Mozurkevich D., Armstrong J. T., Hajian A. R., Elias N. M., 
Hutter D. J., 1998, A J, 116, 2536.  Jennison R. C., 1958, MNRAS, 118, 276. 
 
Kervella P., Traub W. A., Lacasse M. G., 1999, conf. `Working on the Fringe', 
Dana Point, USA, to be published in ASP Conference Series, eds. S. Unwin and R. Stachnik. 
 
Kervella P., Coud\'e du Foresto V., Glindemann A., 2000, Proc. SPIE, conf. 
`Astronomical Interferometry', 4006 (in preparation). 
 
Knox K. T., Thompson B. J., 1974, Ap J, 193, L45. 
 
Koechlin L., Lawson P. R., Mourard D., Blazit A., Bonneau D., Morand F., Stee 
P., Tallon-Bosc I., Vakili F., 1996, Appl. Opt., 35, 3002. 
 
Labeyrie A., 1970, A \& A., 6, 85. 
 
Labeyrie A., 1975, Ap. J., 196, L71. 
 
Labeyrie A., 1978, Ann. Rev. A \& A., 16, 77. 
 
Labeyrie A., 1985, 15th. Advanced Course, Swiss Society of Astrophys. and 
Astron. ed.. A. Benz, M. Huber and M. Mayor, 170. 
 
Labeyrie A., 1995, A \& A, 298, 544. 
 
Labeyrie A., 1996, A \& AS, 118, 517. 
 
Labeyrie A., 1998a, conf. 'Extrasolar planets: 
formation, detection and modeling', Lisbon, Portugal  
(to appear). 
 
Labeyrie A., 1998b,  
Proc. NATO-ASI, conf.`Planets outside the solar system', Carg\`ese, 
Corsica - France. 
 
Labeyrie A., 1998c, Proc. SPIE, conf. `Astronomical interferometry', 3350, 960. 
 
Labeyrie A., 1999, conf. `Working on the fringe', Dana Point, USA, to be 
published in ASP Conference Series, eds. S. Unwin and R. Stachnik. 
 
Labeyrie A., Kibblewhite J., de Graauw T., Roussel Ph., Noordam J., Weigelt G., 
1982b, Proc. CNES, conf. `Very long baseline interferometry', 477. 
 
Labeyrie A., Praderie F., Steinberg J., Vatoux S., Wouters F., 1980, Proc. KPNO, 
conf. `Optical and infrared telescopes for the 1990's', ed. A. Hewitt, 1020. 
 
Labeyrie A., Schumacher G., Savaria E., 1982a, Adv. Space Res., 
2, 11. 
 
Labeyrie A., Schumacher G., Dugu\'e M., Thom C.,  Bourlon P., Foy F., 
Bonneau D. and Foy R., 1986, A \& A., 162, 359. 
 
Lannes A., Anterrieu E., Mar\'echal P., 1997, A \& AS, 123, 183. 
 
Lardi\`ere O., Arnold L., Berger J.-P., Cazal\'e C., Dejonghe J., Labeyrie A., 
Mourard D., 1998, Proc. SPIE, conf. `Telescope Control Systems', 3351, 107. 
 
Lawson P. R., 1994, PASP, 106, 917. 
 
Lawson P. R., 1995, J. Opt. Soc. Am A, 12, 366. 
 
Lawson P. R., Baldwin J. E., Warner P. J., Boysen R. C., Haniff C. A., Rogers 
J., Saint-Jacques D., Wilson D. M. A., Young J. S., 1998, Proc. SPIE, 
conf. `Astronomical Interferometry', 3350, 753. 
 
Lawson P. R., Dumont P. J., Colavita M. M., 1999, AAS Meeting 194. 
 
L\'eger A., Pirre M., Marceau F.J., 1993, A \& A., 277, 309 (1993). 
 
Leinert C., Graser U., 1998, Proc. SPIE, conf. `Astronomical interferometry',  
3350, 389. 
 
L\'ena P., 1997, Experimental Astr., 7, 281. 
 
L\'ena P., Lai O., 1999a, `Adaptive Optics in Astronomy', ed. F. Roddier, 
Cambridge Univ. Press, 351. 
 
L\'ena P., Lai O., 1999b, `Adaptive Optics in Astronomy', ed. F. Roddier, 
Cambridge Univ. Press, 371. 
 
Lindengren L., Perryman M. A. C., 1996, A \& AS, 116, 579. 
 
Linfield R., Gorham P. W., 1999, conf. `Working on the fringe', Dana Point, USA, to 
be published in ASP Conference Series, eds. S. Unwin and R. Stachnik. 
 
Liu Y. C., Lohmann A. W., 1973, Opt. Comm., 8, 372. 
 
Lohmann A. W., Weigelt G. P., Wirnitzer B., 1983, App. Opt., 22, 4028. 
 
Lovelock J.E., 1965, Nature, 207, 568. 
 
Lynds C. R., Worden S. P., Harvey J. W., 1976, Ap J, 207, 174. 
 
Machida Y., Nishikawa J., Sato K., Fukushima T., Yoshizawa M., Honma Y., Torii  
Y., Matsuda K., Kubo K., Ohashi M., Suzuki S., Iwashita H., Proc. SPIE, conf.  
`Astronomical Interferometry', 3350, 202. 
 
Malbet F., Berger J. -P., Colavita M. M., Koresko C. D., Beichman C., Boden A. 
F., Kulkarni S. R., Lane B. F., Mobley D. W., Pan X. -P., Shao M., van Belle G. 
T., Wallace J. K., 1998, astro-ph/9808326, Ap. JL. (accepted). 
 
Mar\'echal P., Anterrieu E., Lannes A., 1997, ASP Conf., 125, 158. 
 
Mariotti J. -M., Menesson B., 1998, Internal ESA report. 
 
Mayor M., Queloz D., Nature, 1995, 378, 355. 
 
McAlister H. A., Bagnuolo W. G., ten Brummelaar, Hartkopf W. I., Shure M. A., 
Sturmann L., Turner N. H., 1998, Proc. SPIE, conf. `Astronomical  
Interferometry', 3350, 947. 
 
McAlister H. A., Bagnuolo W. G., ten Brummelaar, Hartkopf W. I.,  
Turner N. H., Garrison A. K., Robinson W. G., Ridgway S. T., 1994, Proc. SPIE, 
2200, 129.  
 
Mennesson, B., Mariotti J. -M., Coud\'e du Foresto V., Perrin, G., Ridgway S. 
T., Ruilier C., Traub, W. A., Carleton N. P., Lacasse, M. G., Maz\'e G., 1999, 
A \& A, 346, 181. 
 
Mennesson B., Perrin G., Chagnon G., Coud\'e du Foresto V., Morel S., Ruilier 
C., Traub W. A., Carleton N. P., Lacasse M. G., 2000, Proc. SPIE, conf. 
`Interferometry in Optical Astronomy', 4006 (in preparation). 
 
Michelson A. A., 1891, Nature 45, 160. 
 
Michelson A. A., 1920, Ap. J., 51, 257. 
 
Michelson A. A., and Pease F. G., 1921, Ap. J., 53, 249. 
 
Millan-Gabet R., Schloerb P. F., Traub W. A., Carleton N. P., 1999, PASP, 111, 
238. 
 
Millan-Gabet R., Schloerb P. F., Traub W. A., 1998, AAS Meeting 193. 
 
Morel S., Koechlin L., 1998, Proc. SPIE, conf. `Astronomical Interferometry',  
3350, 1057. 
 
Morel S., Traub W. A., Bregman J. D., Mah R., Wilson E., 2000, Proc. SPIE, conf. 
`Interferometry in Optical Astronomy', 4006 (in preparation). 
 
Mourard D., Bonneau D., Koechlin L., Labeyrie A., Morand F., Stee P., 
Tallon-Bosc I., Vakili F., 1997, A \& A, 317, 789. 
 
Mourard D., Bosc I., Labeyrie A., Koechlin A., Saha S., 1989, Nature, 342, 520. 
 
Mourard D., Thureau N., Antonelli P., B\'erio P., Blanc, J.-C., Blazit A., 
Boit J.-L., Bonneau D., Chesneau O., Clausse, J.-M., Corneloup, J.-M., Dalla 
R., Dugu\'e M., Glentzlin A., Hill L., Labeyrie A., Lemerrer J., Menardi S., 
Merlin G., Moreaux, G., Petrov R., Rebattu S., Rousselet-Perraut K., Stee 
P., Tallon-Bosc I., Trastour J., Vakili F.; V\'erinaud C., Voet C., Waultier G., 
1998, Proc. SPIE, conf. `Astronomical Interferometry', 3350, 517. 
 
Mozurkewich D., Johnston K. J., Simon R., Hutter D. J., Colavita M. M., 
Shao M., Pan X.-P., 1991, A J, 101, 2207. 
 
Nakajima T., Kulkarni S. R., Gorham P. W., Ghez A. M., Neugebauer G., Oke 
J. B., Prince T. A., Readhead A. C. S., 1989, A J, 97, 1510. 
 
Ollivier M., Mariotti J.-M., 1997, Appl. Opt., 36, 5340. 
 
Padilla C. E., Karlov V. I., Matson L. K., Soosaar K., Brummelaar T. ten, 
1998, Proc. SPIE, conf. `Astronomical Interferometry', 3350, 1045. 
 
Pan X.-P., Shao M., Colavita M. M., 1992, IAU Colloq. 135., ASP Conf. Proc. 32, 
`Complementary Approaches to Double and Multiple Star Research', eds. H. A. 
McAlister and W. I. Hartkopf, 502. 
 
Pan X.-P., Kulkarni S. R., Colavita M. M., Shao M., 1996, Bull. Am. Astron. 
Soc., 28, 1312. Pauls T. A., Mozurkewich D., Armstrong J. T., Hummel C. A.,  
Benson J. A., Hajian A. R., 1998, Proc. SPIE, conf. `Astronomical  
Interferometry', 3350, 467. 
 
Pedretti E., Labeyrie A., 1999, A \& AS, 137, 543. 
 
Penny A. J., L\'eger A., Mariotti J. -M., Schalinski C., Eiora C., Laurance R., 
Fridlund M., 1998, Proc. SPIE, conf. `Astronomical Interferometry', 3350, 666. 
 
Perrin G., 1997, A \& AS, 121, 553. 
 
Perrin G., Coud\'e du Foresto V., Ridgway S. T., Mariotti J.-M., Traub W. A., 
Carleton N. P., Lacasse M. G., 1998, A \& A, 331, 619. 
 
Perrin G., Coud\'e du Foresto V., Ridgway S. T., Menesson B., Ruilier C., 
Mariotti J -M., Traub W. A., Lacasse M. G., 1999, A \& A, 345, 221. 
 
Perryman M. A. C., 1998, Nature, 340, 111. 
 
Petrov R., Roddier F., Aime C., 1986, J. Opt. Soc. Am. A, 3, 634. 
 
Petrov R., Malbet F., Richichi A., Hofmann K. H., Agabi K., Antonelli P., 
Aristidi E., Baffa C., Beckmann U., B\'erio  P., Bresson Y., Cassaing  F., 
Chelli A., Dress A., Dugu\'e M., Duvert G., Forveille T., Fossat E., Gennari 
S., Geng M., Glentzlin A., Kamm D., Lagarde S., Lecoarer E., Le Contel 
J.-M., Lisi F., Lopez B., Mars G., Martinot-Lagarde G., Monin J., Mouillet 
D., Mourard D., Rousselet-Perraut K., Perrier-Bellet C., Puget P., Rabbia Y., 
Rebattu S., Reynaud F., Robbe-Dubois S., Sacchettini M., I. Tallon-Bosc, 
Weigelt  G., 2000, Proc. SPIE, conf. `Interferometry in Optical Astronomy',  
4006 (in preparation). 
 
Quirrenbach A., Coud\'e du Foresto V., Daigne G., Hofmann K. H., 
Hofmann R., Lattanzi M., Osterbart R., Le Poole R. S., Queloz D., Vakili F., 
1998, Proc. SPIE, conf. `Astronomical Interferometry', 3350, 807. 
 
Rabbia Y., Mekarnia D., Gay J., 1990, Proc. SPIE, conf. `Infrared Technology',  
1341, 172. 
 
Rhodes W. T., Goodman J. W., 1973, J. Opt. Soc. Am., 63, 647. 
 
Roddier C., Roddier F., 1988, Proc. NATO-ASI, conf. `Diffraction Limited Imaging 
with Very Large Telescopes', eds. D. M. Alloin and J. -M. Mariotti, Carg\`ese,  
Corsica - France, 221. 
 
Rousset G., Fontanella J. C., Kem P., Gigan P., Rigaut F., L\'ena P., 
Boyer P., Jagourel P., Gaffard J. P., Merkle F., 1990, A \& A, 230, L29. 
 
Rousselet-Perraut K. , Vakili F. , Mourard D., 1996, Opt. Engin., 35, 2943. 
 
Saha S. K., 1999a, BASI., 27, 443. 
 
Saha S. K., 1999b, Ind. J. Phys., 73B, 552. 
 
Sato K., Nishikawa J., Yoshizawa M., Fukushima T., Machida Y., Honma Y.,  
Kuwabara R., Suzuki S., Torii Y., Kubo K., Matsuda K., Iwashita H., Proc. SPIE, 
conf. `Astronomical Interferometry', 3350, 212. 
 
Schloerb F. P., Millan-Gabet, R. S., Traub, W. A., 1999, AAS Meeting 194. 
 
Serabyn E., 1999, Appl. Opt., 38, 4213. 
 
Shaklan S. B., Roddier F., 1987, Appl. Opt., 26, 2159. 
 
Shao M., Colavita M. M., 1988, A \& A, 193, 357. 
 
Shao M., Colavita M. M., 1994, IAU Symp. 158, `Very high resolution imaging', 
eds. J. G. Robertson and W. J. Tango, 413. 
 
Shao M., Colavita M. M., Hines B. E., Hershey J. L., Hughes J. A., Hutter 
D. J., Kaplan G. H., Johnston K. J., Mozurkewich D., Simon R. H., Pan X. -P.,  
1990, A J, 100, 1701. 
 
Shao M., Colavita M. M., Hines B. E., Staelin D. H., Hutter D. J., Johnston K. 
J., Mozurkewich D., Simon R. H., Hershey J. L., Hughes J. A., Kaplan G. H., 
1988, A \& A, 193, 357. 
 
Shao M., Staelin D. H., 1977, J. Opt. Soc. Am., 67, 81. 
 
Stee P., de Ara\'ujo, Vakili F., Mourard D., Arnold I., Bonneau D., Morand F., 
Tallon-Bosc I., 1995, A \& A, 300, 219. 
 
Stee P., Vakili F., Bonneau D., Mourard D., 1998, A \& A, 332, 268. 
 
St\'ephan H., 1874, C. R. Acad. Sci. Paris, 76, 1008. 
 
Tatarski V. I., 1967, `Wave Propagation in a Turbulent Medium', Dover, N. Y. 
 
Thom C., Granes P., Vakili F., 1986, A \& A, 165, L13. 
 
Traub W. A., Millan-Gabet R., Garcia M. R., 1998, AAS Meeting 193. 
 
Traub W. A., Carleton N. P., Brewer M. K., Lacasse M. G., Millan-Gabet R., 
Morel S., Papaliolios C., Porro I., 2000, Proc. SPIE, conf. `Interferometry in  
Optical Astronomy', 4006, (in preparation). 
 
Unwin S. C., Turyshev S. G., Shao M., 1998, Proc. SPIE, conf. `Astronomical 
Interferometry', 3350, 551. 
 
Unwin S. C., 1999, Private communication. 
 
Vakili F., B\'erio P., Bonneau D., Chesneau O., Mourard D., Stee P., Thureau N., 
1998a, conf. `Be stars', ed. A. M. Hubert and C. Jaschek, 173. 
 
Vakili F., Mourard D., Bonneau D., Morand F., Stee P., 1997, A \& A, 323, 183. 
 
Vakili F., Mourard D., Stee P., Bonneau D., B\'erio P., Chesneau O., Thureau N., 
Morand F., Labeyrie A., Tallon-Bosc I., 1998b, A \& A, 335, 261. 
 
Van Belle G. T., Dyck H. M., Benson J. A., Lacasse M. G., 1996, A J., 112, 
2147. 
 
Van Belle G. T., Dyck H. M., Thompson R. R., Benson J. A., Kannappan S. J., 
1997, A J., 114, 2150. 
 
Van Belle G. T., Lane B. F., Thompson R. R., Boden A. F., Colavita M. M., 
Dumont P. J., Mobley D. W., Palmer D;, Shao M., Vasisht G. X., Wallace J. K., 
Creech-Eakman M. J., Koresko C. D., Kulkarni S. R., Pan X.-P., Gubler J., 
1999, A J, 117, 521. 
 
V\'erinaud C., Blazit A., de Bonnevie A., B\'erio P., 1998, conf. 
`Catching the Perfect Wave', eds. S. R. Restaino, W. Junor and N. Duric., 131. 
 
Walkup J. F., Goodman J. W., 1973, J. Opt. Soc. Am., 63, 399. 
 
Wallace J. K., Boden A. F., Colavita M. M., Dumont P. J., Gursel Y., Hines B.,  
Koresko C. D., Kulkarni S. R., Lane B. F., Malbet F., Palmer D., Pan X. P.,  
Shao M., Vasisht G. X., Van Belle G. T.,  
Yu J., 1998, Proc. SPIE, conf. `Astronomical interferometry', 3350, 864. 
 
Weigelt G., 1977, Opt. Communication, 21, 55. 
 
Weigelt G., Mourard D., Abe L., Beckmann U., Bl\"oecker T., Chesneau O.,  
Hillemanns C., Hoffmann K. H., Ragland S., Schertl D., Scholz M., Stee P.,  
Thureau N., Vakili F., 2000, Proc. SPIE, conf. `Interferometry in Optical  
Astronomy', 4006, (in preparation). 
}                                         
 
\end